\def\IEEEsubmission{0}
\def\reviewColor{black}
\def\figuresize{3.3in}

\def\complexNumbers{\mathbb{C}}

\def\functionSpace[#1]{\mathcal{F}(#1)}
\def\realNumbers{\mathbb{R}}
\def\integers{\mathbb{Z}}
\def\constante{{\rm e}}
\def\constantj{{\rm j}}
\def\expectationOperator[#1][#2]{\mathbb{E}_{#2}\left[#1\right]}
\def\uniformDistribution[#1][#2]{{\mathcal{U}_{[#1,#2]}}}
\def\traceOperator[#1]{{\mathrm{tr}}\{#1\}}
\def\identityMatrix[#1]{\mathrm{\textbf{I}}_{#1}}
\def\zeroVector[#1]{{ {{\mathbf{0}}}}_{#1}}
\def\oneVector[#1]{{ {\mathbf{1}}}_{#1}}
\def\exponentialIntegral[#1]{\mathrm{Ei}(#1)}
\def\functionArbitrary[#1]{f_{#1}}
\def\functionArbitraryEstimate[#1]{\hat{f}_{#1}}
\def\indicatorFunction[#1]{\mathbb{I}\left[{#1}\right]}

\def\clamp[#1][#2]{\text{clamp}_{#2}\left(#1\right)}
\def\signNormal[#1]{\text{sign}\left(#1\right)}
\def\diagOperation[#1]{\text{diag}\left\{#1\right\}}

\def\probability[#1]{\mathrm{Pr}({#1})}
\def\complexGaussian[#1][#2]{\mathcal{CN}({#1,#2})}
\def\gaussian[#1][#2]{\mathcal{N}({#1,#2})}

\def\normalPDF[#1]{\phi\left(#1\right)}

\def\indexSample{p}
\def\samplePoint[#1][#2]{y_{#1}^{(#2)}}
\def\CDF[#1][#2][#3]{F_{#1}^{#2}\left({#3}\right)}

\def\PDF[#1][#2][#3]{f_{#1}^{#2}\left({#3}\right)}

\def\apac[#1][#2]{a_#2}
\def\apacPositive[#1][#2]{\rho^{+}_{#1}(#2)}
\def\lagForCorrelation{\ell}
\def\indexEleOfSeq{n}

\def\seqGx{\textit{\textbf{x}}}
\def\eleGx[#1]{{x}_{#1}}
\def\eleGxConj[#1]{\overline{x}_{#1}}
\def\seqGxP{X}
\def\seqGaP{A}
\def\rootSeqX[#1]{\alpha_{#1}}
\def\rootSeqXConj[#1]{\overline{\alpha}_{#1}}

\def\polySeq[#1][#2]{{#1}(#2)}
\def\polySeqConj[#1][#2]{\overline{{#1}(#2)}}
\def\polyVariable{z}

\def\indexRoot{k}
\def\indexRootRX{l}
\def\indexRootRXBin[#1]{l_{#1}}
\def\indexRootDigit{i}

\def\numberOfMVs{M}
\def\numberOfEdgeDevices{U}
\def\numberOfEdgeDevicesPlus[#1]{U^{+}_{#1}}
\def\numberOfEdgeDevicesMinus[#1]{U^{-}_{#1}}

\def\indexED{u}
\def\indexTime{n}

\def\funcSignalOne[#1]{\mathcal{X}_{1}(#1)}
\def\funcSignalTwo[#1]{\mathcal{X}_{2}(#1)}
\def\funcSignalThree[#1]{\mathcal{X}_{3}(#1)}
\def\funcChannel[#1]{\Gamma(#1)}
\def\funcNoise[#1]{{\Omega}(#1)}

\def\pdpTap[#1]{\rho_{#1}}
\def\decayingParameter{\rho}

\def\computationErrorRate[#1]{{\tt{CER}}_{#1}}
\def\probabilityErrorPlus[#1]{{\tt{CER}^{+}}}
\def\rate[#1]{\lambda_{#1}}

\def\voteVectorEDBinaryEle[#1][#2]{b^{(#1)}_{#2}}
\def\voteVectorEDEle[#1][#2]{v^{(#1)}_{#2}}
\def\voteVectorED[#1]{{\textit{\textbf{v}}}^{(#1)}}
\def\voteVectorAcrossED[#1]{\textit{\textbf{u}}_{#1}}
\def\voteAll{\textit{{\textbf{V}}}}
\def\voteAllWithout{\dot{\textit{{\textbf{V}}}}}

\def\majorityVoteEle[#1]{m_{#1}}
\def\majorityVoteDetectedEle[#1]{\hat{m}_{#1}}

\def\indexMV{\ell}

\def\metricPlus[#1]{\tilde{U}^{+}_{#1}}
\def\metricMinus[#1]{\tilde{U}^{-}_{#1}}

\def\coefOneFcn[#1][#2]{a^{#2}{(#1)}}

\def\codedSeqP[#1]{X^{(#1)}}
\def\codedSeq[#1]{\textit{\textbf{x}}^{(#1)}}
\def\codedSeqEle[#1][#2]{{x}^{(#1)}_{#2}}
\def\codedSeqEleConj[#1][#2]{{\bar{x}}^{(#1)}_{#2}}
\def\rootCodedSeq[#1][#2]{\alpha^{(#1)}_{#2}}
\def\rootCodedSeqConj[#1][#2]{\bar{\alpha}^{(#1)}_{#2}}
\def\numberOfRoots{K}
\def\indexRoot{k}
\def\radiusRoot{d}
\def\setHuffmanRoots[#1]{\mathcal{Z}_{#1}}
\def\lastCorrTerm{\eta}

\def\transmittedSeqP[#1]{T^{(#1)}}
\def\transmittedSeq[#1]{\textit{\textbf{t}}^{(#1)}}
\def\transmittedSeqEle[#1][#2]{{t}^{(#1)}_{#2}}

\def\channelSeqP[#1]{H^{(#1)}}
\def\channelSeq[#1]{\textit{\textbf{h}}^{(#1)}}
\def\channelSeqEle[#1][#2]{h^{(#1)}_{#2}}
\def\rootChannelSeq[#1][#2]{\gamma^{(#1)}_{#2}}
\def\channelLength{L_{\rm e}}
\def\indexTap{l}

\def\integralVar{t}
\def\CDFvariable{x}
\def\charFcn[#1]{\varphi(#1)}
\def\charFcnPlus[#1]{\varphi^{+}_{\indexMV}(#1)}
\def\charFcnMinus[#1]{\varphi^{-}_{\indexMV}(#1)}
\def\charFcnMinusConj[#1]{{{\varphi}^{-}_{\indexMV}(#1)}^*}

\def\metricOne{A}
\def\metricTwo{B}
\def\valOne[#1]{a_{#1}}
\def\valTwo[#1]{b_{#1}}
\def\rateOne[#1]{\lambda_{a_{#1}}}
\def\rateTwo[#1]{\lambda_{b_{#1}}}
\def\charFcnOne[#1]{\varphi_{a}(#1)}
\def\charFcnTwo[#1]{\varphi_{b}(#1)}
\def\charFcnTwoConj[#1]{{{\varphi}_{b}^*(#1)}}

\def\superposedSeqP{S}

\def\receivedSeqP{R}
\def\receivedSeq[#1]{\textit{\textbf{r}}_{#1}}
\def\receivedSeqEle[#1]{{r}_{#1}}

\def\noiseSeqP{W}
\def\noiseSeq{\textit{\textbf{w}}}
\def\noiseSeqEle[#1]{{w}_{#1}}

\def\SNR{{\tt{SNR}}}

\def\noiseSample[#1]{\omega_{#1}}
\def\transmitPower[#1]{P_{#1}}
\def\noiseVariance{\sigma_{\rm noise}^2}

\def\goldenbaumLength{L_{\rm seq}}

\if\IEEEsubmission1
\documentclass[journal,12pt,onecolumn,draftclsnofoot]{IEEEtran}
\else
\documentclass[journal]{IEEEtran}
\fi
\usepackage{multicol}
\usepackage{lipsum}
\usepackage{mathtools}
\usepackage{amsthm}
\usepackage{acronym}
\usepackage[bookmarksopen=true]{hyperref}
\usepackage{stfloats}
\usepackage{amsfonts}
\usepackage{acronym}
\usepackage[noadjust]{cite}
\usepackage{multirow}
\usepackage{bm}
\usepackage{amsmath,amssymb}
\usepackage{graphicx}
\usepackage{epstopdf}
\usepackage[table,xcdraw]{xcolor}
\usepackage[caption=false,font=footnotesize]{subfig}
\usepackage[geometry]{ifsym}
\usepackage{array}
\usepackage[utf8]{inputenc}
\usepackage[T1]{fontenc}
\usepackage{pifont}
\usepackage{tabularray}
\usepackage{lipsum}
\def\BibTeX{{\rm B\kern-.05em{\sc i\kern-.025em b}\kern-.08em
		T\kern-.1667em\lower.7ex\hbox{E}\kern-.125emX}}

\let\norm\undefined % <-- "Undefine" \norm
\DeclarePairedDelimiter\norm{\lVert}{\rVert}
\newcommand\mydots{\hbox to 1em{.\hss.\hss.}}

\DeclarePairedDelimiter\floor{\lfloor}{\rfloor}

\newtheorem{definition}{Definition}
\newtheorem{lemma}{Lemma}

\newtheorem{corollary}{Corollary}
 
\newtheorem{example}{\color{black} Example}

\DeclareMathOperator{\sign}{sign}

\makeatletter
\newif\ifAC@uppercase@first%
\def\Aclp#1{\AC@uppercase@firsttrue\aclp{#1}\AC@uppercase@firstfalse}%
\def\AC@aclp#1{%
	\ifcsname fn@#1@PL\endcsname%
	\ifAC@uppercase@first%
	\expandafter\expandafter\expandafter\MakeUppercase\csname fn@#1@PL\endcsname%
	\else%
	\csname fn@#1@PL\endcsname%
	\fi%
	\else%
	\AC@acl{#1}s%
	\fi%
}%
\def\Acp#1{\AC@uppercase@firsttrue\acp{#1}\AC@uppercase@firstfalse}%
\def\AC@acp#1{%
	\ifcsname fn@#1@PL\endcsname%
	\ifAC@uppercase@first%
	\expandafter\expandafter\expandafter\MakeUppercase\csname fn@#1@PL\endcsname%
	\else%
	\csname fn@#1@PL\endcsname%
	\fi%
	\else%
	\AC@ac{#1}s%
	\fi%
}%
\def\Acfp#1{\AC@uppercase@firsttrue\acfp{#1}\AC@uppercase@firstfalse}%
\def\AC@acfp#1{%
	\ifcsname fn@#1@PL\endcsname%
	\ifAC@uppercase@first%
	\expandafter\expandafter\expandafter\MakeUppercase\csname fn@#1@PL\endcsname%
	\else%
	\csname fn@#1@PL\endcsname%
	\fi%
	\else%
	\AC@acf{#1}s%
	\fi%
}%
\def\Acsp#1{\AC@uppercase@firsttrue\acsp{#1}\AC@uppercase@firstfalse}%
\def\AC@acsp#1{%
	\ifcsname fn@#1@PL\endcsname%
	\ifAC@uppercase@first%
	\expandafter\expandafter\expandafter\MakeUppercase\csname fn@#1@PL\endcsname%
	\else%
	\csname fn@#1@PL\endcsname%
	\fi%
	\else%
	\AC@acs{#1}s%
	\fi%
}%
\edef\AC@uppercase@write{\string\ifAC@uppercase@first\string\expandafter\string\MakeUppercase\string\fi\space}%
\def\AC@acrodef#1[#2]#3{%
	\@bsphack%
	\protected@write\@auxout{}{%
		\string\newacro{#1}[#2]{\AC@uppercase@write #3}%
	}\@esphack%
}%
\def\Acl#1{\AC@uppercase@firsttrue\acl{#1}\AC@uppercase@firstfalse}
\def\Acf#1{\AC@uppercase@firsttrue\acf{#1}\AC@uppercase@firstfalse}
\def\Ac#1{\AC@uppercase@firsttrue\ac{#1}\AC@uppercase@firstfalse}
\def\Acs#1{\AC@uppercase@firsttrue\acs{#1}\AC@uppercase@firstfalse}

\acrodef{WSN}{wireless sensor network}
\acrodef{USRP}{universal software radio peripheral}
\acrodef{SN}{sensor node}
\acrodef{FC}{fusion center}
\acrodef{MAC}{multiple-access channel}
\acrodef{FL}{federated learning}
\acrodef{ED}{edge device}
\acrodef{CS}{compressed sensing}
\acrodef{ES}[BS]{base station}
\acrodef{DCN}{data center network}
\acrodef{RIS}{reconfigurable intelligent surfaces}
\acrodef{IMC}{in-memory computing}
\acrodef{FPGA}{field-programmable gate array}
\acrodef{SDR}{software-defined radio}
\acrodef{PS}{processing system}
\acrodef{SS}{soft synchronization}
\acrodef{IQ}{in-phase/quadrature}
\acrodef{IP}{intellectual property}
\acrodef{DMA}{direct-memory access}
\acrodef{RAM}{random access memory}
\acrodef{CC}{companion computer}
\acrodef{FEE}{function estimation error}
\acrodef{MSK}{minimum-shift keying}
\acrodef{TDMA}{time-domain multiple access}
\acrodef{PLNC}{physical-layer network coding}
\acrodef{UAV}{unmanned aerial vehicle}
\acrodef{LoRa}{Long-Range}
\acrodef{DC}{direct-current}
\acrodef{DAC}{digital-to-analog converter}
\acrodef{ADC}{anlog-to-digital converter}
\acrodef{CS}{complementary sequence}
\acrodef{GCP}{Golay complementary pair}
\acrodef{ANF}{algebraic normal form}
\acrodef{AACF}{aperiodic auto-correlation function}
\acrodef{RM}{Reed-Muller}
\acrodef{MOCZ}{modulation on conjugate-reciprocal zeros}
\acrodef{BMOCZ}{binary modulation on conjugate-reciprocal zeros}
\acrodef{dizet}[DiZeT]{direct zero-testing}

\acrodef{PUCCH}{physical uplink control channel}
\acrodef{PRACH}{physical random access channel}

\acrodef{OBO}{output-power back-off}
\acrodef{ACLR}{adjacent-channel-leakage ratio}

\acrodef{LDPC}{low-density parity check}

\acrodef{PDF}{probability density function}
\acrodef{CDF}{cumulative distribution function}

\acrodef{TBMA}{type-based multiple access}

\acrodef{MSFE}{mean-squared function error}
\acrodef{FEE}{function-estimation error}
\acrodef{CER}{computation error rate}
\acrodef{BCER}{block-computation error rate}
\acrodef{CFO}{carrier frequency offset}
\acrodef{TO}{time offset}
\acrodef{PO}{phase offset}
\acrodef{RSSI}{received signal strength  information}

\acrodef{STLC}{space-time line code}
\acrodef{CCI}{co-channel interference}
\acrodef{CSIT}[CSIT]{\ac{CSI} at the transmitter}
\acrodef{CSIR}[CSIR]{\ac{CSI} at the receiver}
\acrodef{MIMO}{multiple-input-multiple-output}
\acrodef{PC}{phase correction}
\acrodef{ZF}{zero-forcing}
\acrodef{ANOVA}{analysis of variance}

\acrodef{PCA}{principal component analysis}
\acrodef{TIG}{Technical Interest Group}

\acrodef{FSK}{frequency-shift keying}
\acrodef{PPM}{pulse-position modulation}
\acrodef{PAM}{pulse-amplitude modulation}

\acrodef{MRC}{maximum-ratio combining}
\acrodef{HP}{hard-coded participation}
\acrodef{HPA}{hard-coded participation with absentees}
\acrodef{SP}{soft-coded participation}
\acrodef{FSK-MV}{\ac{FSK}-based \ac{MV}}
\acrodef{RF}{radio-frequency}
\acrodef{MF}{matched filter}
\acrodef{PPM}{pulse-position modulation}
\acrodef{CSK}{chirp-shift keying}
\acrodef{PPM-MV}[PPM-MV]{\ac{PPM}-based \ac{MV}}
\acrodef{DFT-s-OFDM}{discrete Fourier transform-spread orthogonal frequency division multiplexing}
\acrodef{SC}{single-carrier}
\acrodef{SGD}{stochastic gradient descent}
\acrodef{signSGD}{sign stochastic gradient descent}

\acrodef{SL}{split learning}
\acrodef{SNR}{signal-to-noise ratio}
\acrodef{RMSE}{root-mean-squared error}
\acrodef{OFDM}{orthogonal frequency division multiplexing}
\acrodef{DFT}{discrete Fourier transform}
\acrodef{PSK}{phase-shift keying}
\acrodef{QAM}{quadrature amplitude modulation}
\acrodef{QPSK}{quadrature phase-shift keying}
\acrodef{PMEPR}{peak-to-mean envelope power ratio}
\acrodef{BER}{bit-error ratio}
\acrodef{SNR}{signal-to-noise ratio}
\acrodef{PSD}{power spectral density}
\acrodef{SE}{spectral efficiency}
\acrodef{CP}{cyclic prefix}
\acrodef{AWGN}{additive white Gaussian noise}
\acrodef{CFR}{channel frequency response}
\acrodef{CIR}{channel impulse response}
\acrodef{MMSE}{minimum mean-squared error}
\acrodef{LMMSE}{linear minimum mean-squared error}
\acrodef{BPSK}{binary phase shift keying}
\acrodef{QPSK}{quadrature phase shift keying}
\acrodef{BLER}{block-error rate}
\acrodef{ML}{maximum likelihood}
\acrodef{PHY}{physical layer}
\acrodef{PA}{power amplifier}
\acrodef{IDFT}{inverse discrete Fourier transform}
\acrodef{DoF}{degrees-of-freedom}
\acrodef{IoT}{Internet-of-Things}
\acrodef{FDE}{frequency-domain equalization}
\acrodef{RF}{radio-frequency}
\acrodef{IM}{index modulation}
\acrodef{MF}{matched filter}
\acrodef{PPM}{pulse-position modulation}

\acrodef{MSE}{mean-squared error}
\acrodef{MRT}{maximum-ratio transmission}
\acrodef{ERC}{equal-ratio combining}
\acrodef{BAA}{broadband analog aggregation}
\acrodef{OBDA}{one-bit broadband digital aggregation}
\acrodef{FEEL}{federated edge learning}
\acrodef{FL}{federated learning}
\acrodef{UL}{uplink}
\acrodef{DL}{downlink}
\acrodef{OAC}{over-the-air computation}
\acrodef{TCI}{truncated-channel inversion}
\acrodef{MV}{majority vote}
\acrodef{CNN}{convolution neural network}
\acrodef{ReLU}{rectified-linear unit}
\acrodef{CSI}{channel state information}
\acrodef{PAPR}{peak-to-average power ratio}
\acrodef{SC}{single-carrier}
\acrodef{iid}[IID]{independent and identically distributed}
\acrodef{RMS}{root-mean-square}
\acrodef{4G}{Fourth Generation}
\acrodef{5G}{Fifth Generation}
\acrodef{NR}{New Radio}
\acrodef{LTE}{Long-Term Evolution}
\acrodef{OFDMA}{orthogonal frequency division multiple access}
\acrodef{HARQ}{hybrid automatic repeat request}
\acrodef{D2D}{Device-to-Device}
\acrodef{NOMA}{non-orthogonal multiple access}
\acrodef{OMA}{orthogonal multiple access}

\acrodef{IMT}{International Mobile Telecommunications}
\acrodef{ITU}{International Telecommunication Union}

\acrodef{PDP}{power-delay profile}
\acrodef{TBMA}{type-based multiple access}
\begin{document}

\title{
	{Over-the-Air Majority Vote Computation with Modulation on Conjugate-Reciprocal Zeros}\\
	%{On Reliable Over-the-Air Computation of Majority Vote Function with Complementary Sequences}\\
	\thanks{Alphan~\c{S}ahin is with the Electrical  Engineering Department,
		University of South Carolina, Columbia, SC, USA. E-mail: asahin@mailbox.sc.edu}
		\thanks{This paper was submitted in part to the IEEE Globel Communications Conference 2024  \cite{sahinGC_2024}.}	
	\author{Alphan~\c{S}ahin,~\IEEEmembership{Member,~IEEE}} 
}
\maketitle

\begin{abstract}
In this study, we propose a new approach to compute the \ac{MV} function based on \ac{MOCZ} and introduce three different methods. 
%The proposed methods rely on the fact that when a linear combination of polynomials is evaluated at one of the roots of a polynomial in the combination, that polynomial does contribute to the evaluation.
%The proposed methods rely  on evaluating the polynmials at one of the roots of a polynomial in the combination, that polynomial does contribute to the evaluation.
%Using the zeros of polynomials for OAC is challenging since the signal superposition changes the original zeros of the polynomials at the transmitters non-linearly. 
%To utilize this property, 
In these methods, each transmitter maps the votes to the zeros of a Huffman polynomial, and the corresponding polynomial coefficients are transmitted. The receiver evaluates the polynomial constructed by the elements of the superposed sequence at conjugate-reciprocal zero pairs and detects the MV with a \ac{dizet} decoder. With differential and index-based encoders, we eliminate the need for power-delay information at the receiver while improving the \ac{CER} performance.
The proposed methods do not use instantaneous channel state information at the transmitters and receiver. Thus, they provide robustness against phase and time synchronization errors. We theoretically analyze the \acp{CER} of the proposed methods. Finally, we demonstrate their efficacy in a distributed median computation scenario.
% We show that the \ac{CER} can be reduced by using  in fading channels.
\end{abstract}
\section{Introduction}
\acresetall

{\color{\reviewColor}With more applications relying on data exchange across many distributed nodes over wireless networks, meeting the time requirements under limited communication bandwidth becomes increasingly challenging. One of the prominent approaches to reduce the latency for these scenarios is \ac{OAC}
 %by exploiting the signal superposition property of wireless multiple-access channels
 \cite{Nazer_2007,goldenbaum2013harnessing,goldenbaum2015nomographic}. 
With OAC, 
%the nodes transmit their readouts on the same spectral resources concurrently, and naturally occurring interference in the wireless channel realizes the underlying sum function for evaluating a desired multivariate function.
%a multivariate function is aimed to be computed through signal superposition via simultaneous transmissions along with proper encoders and a decoder.
a multivariate function is aimed to be computed by relying on its representation that can structurally match with the underlying operation that multiple-access wireless channel naturally performs.
%the information from each node is not acquired independently with OAC, 
%naturally occurring 
The remarkable gain obtained with OAC is that latency does not scale with the number of nodes since multi-user interference is harnessed for computation  through simultaneous transmissions
\cite{sahinSurvey2023}. OAC has been considered for a wide range of applications, such as wireless federated learning \cite{zhu2021over}, distributed localization \cite{safiMVlocOJCS}, wireless data centers \cite{Xiugang_2016}, and wireless control systems \cite{Cai_2018}.  However, achieving a reliable computation with OAC is difficult because signal superposition occurs after the channel distorts the signals. In this work, we focus on this fundamental problem and propose an \ac{OAC} scheme that does not rely on the availability of \ac{CSI} at the transmitters and receiver, based on a recently proposed method, i.e., {\em \ac{MOCZ}} \cite{Walk_2017ISIT,Philipp_2019principles,Philipp_2020practical}. 
%We introduce several methods to compute the \ac{MV} function. In following section, we provide further details on \ac{MOCZ}, \ac{OAC}, and the applications of \ac{MV} function. 
}

%In practice, achieving a reliable computation with OAC is a difficult problem because the signal superposition occurs after the wireless channel distorts the signals.
{\color{\reviewColor}In the literature, there is a substantial effort to overcome fading channels for OAC \cite{Altun_2021survey,sahinSurvey2023,Zheng_2023,hellstrom2020wireless,Zhibin_2022oac}.} A commonly used technique is channel inversion at the transmitter, where the transmitter multiplies the parameters with the inverses of the channel coefficients before transmission \cite{Amiri_2020,Guangxu_2020,Guangxu_2021,Wei_2022,Yao_2023}. {\color{\reviewColor}While this solution ensures coherent signal superposition, it is sensitive to phase errors at the transmitters, as the receiver cannot fix the potential phase errors with linear equalizers.}
Phase synchronization can be very challenging in practice because of the inevitable hardware impairments such as clock errors, residual \ac{CFO}, calibration errors, and jittery time synchronization. For instance,  a large phase rotation in the frequency domain can occur because of sample deviations at the transmitter or receiver  \cite{sahinGC_2022}. Also, non-stationary  channel conditions can deteriorate the coherent signal superposition in mobile environments \cite{Haejoon_2021}. {\color{\reviewColor}To eliminate the need for phase synchronization,
 non-coherent OAC schemes that exploit the energy of the superposed signals have been studied in the literature.
% one solution is to use non-coherent superposition.
For instance, in \ac{TBMA}, the nodes transmit on orthogonal resources allocated for different classes (i.e., types),  and the frequency histogram is estimated to compute statistical averages \cite{Mergen_2006tsp,Mergen_2007}.
In \cite{Sahin_2022MVjournal} and \cite{safiMVlocOJCS}, two orthogonal resources representing the local gradient directions (i.e., $1$ and $-1$ as votes) are allocated, and the norms of received symbols on these resources are compared with each other to compute the \ac{MV} for the global gradient direction. 
Similarly, in \cite{sahin2022md}, the digits of the parameters are proposed to be aggregated on orthogonal resources dedicated to all possible numerals in a balanced number system. An alternative non-coherent OAC solution is  Goldenbaum's scheme  \cite{Goldenbaum_2013tcom,Goldenbaum_wcl2014}.} In this method, a random unimodular sequence is multiplied by the square root of the parameter to be aggregated. At the receiver, the norm-square of the received sequence is calculated to estimate the aggregated parameter. Although this method can provide robustness against synchronization errors, it can suffer from interference terms due to the loss of orthogonality between the sequences. 
{\color{\reviewColor}It is also worth noting that non-coherent OAC schemes demonstrably work in practice without introducing stringent requirements. For instance, in \cite{sahinGC_2022}, the distributed training based on \ac{MV} computation is demonstrated with off-the-shelf \acp{SDR}.  Goldenbaum's scheme is also shown in practice in \cite{Kortke_2014}.}  %Given its robustness against phase synchronization errors, we consider non-coherent OAC in this work.}

{\color{\reviewColor}With \Ac{OAC}, a set of  special functions, i.e., nomographic functions, can be computed. Several examples of nomographic functions are arithmetic mean,  weighted sum,  modulo-2 sum, polynomial function, maximum, product, and \ac{MV} \cite{sahinSurvey2023}. In this work, we are interested in computing the \ac{MV} function with OAC. \ac{MV} function is utilized for a wide variety of applications. For example, \cite{Guangxu_2021,Sahin_2022MVjournal}, and \cite{mohammadICC_2022} implement distributed training based on \ac{MV} \cite{Bernstein_2018} over wireless networks with OAC.  In \cite{safiMVlocOJCS}, the \ac{MV} function is employed for distributed localization with the motivation of enhanced security. In \cite{sahin2023reliable}, \ac{MV} is considered for \ac{UAV} waypoint flight control. Another appealing feature of the \ac{MV} function is that it can be used for median computation, as discussed further in Section~\ref{sec:numerical}. Compared to the arithmetic mean, aggregating parameters based on the median can provide immunity against outliers or Byzantine attacks \cite{Chennips_2020,Miao_2024}.} 
%We also consider the median computation in distributed scenario, , to demonstrate the performance of the proposed OAC scheme.}

\ac{MOCZ} is a non-linear modulation technique where information bits are encoded into the zeros of a polynomial, and the transmitted sequence corresponds to the polynomial coefficients \cite{Walk_2017ISIT}. As comprehensively analyzed in  \cite{Philipp_2019principles}, the merit of \ac{MOCZ} is that the zero structure of the transmitted signal is preserved at the receiver regardless of \ac{CIR}. This property is because the convolution operation can be represented as a polynomial multiplication, and the zeros are unaffected by the multiplication operation. As a result, \ac{MOCZ} enables the receiver to obtain the information bits  without the knowledge of instantaneous \ac{CSI}. Although the idea of modulation on zeros can be used with an arbitrary set of polynomials, the zeros of a polynomial can be very sensitive to perturbation of its coefficients (e.g., Wilkinson's polynomial \cite{Wilkinson_1984}). {\color{\reviewColor}To achieve robustness against additive noise, in \cite{Philipp_2019principles}, the authors propose to use Huffman polynomials \cite{Huffman_1962},  discussed further in Section~\ref{sec:system}.} In the literature, \ac{MOCZ} is evaluated and improved in various scenarios. For instance, in \cite{Philipp_2020practical}, the authors investigate the practical aspects of \ac{MOCZ} and assess its performance under impairments like carrier frequency and time offsets. In \cite{Philipp_2021mumocz}, \ac{MOCZ} is investigated along with \ac{DFT-s-OFDM} and extended to multi-user scenarios. In \cite{dehkordi2023integrated}, the correlation properties of Huffman sequences are exploited to achieve joint radar and communications with \ac{MOCZ} at 60~GHz millimeter wave band. In \cite{Yaping_2023}, the authors consider multiple antennas and develop a non-coherent Viterbi-like detector to achieve diversity gain. In \cite{Siddiqui_2024}, the authors consider an over-complete system for MOCZ based on  faster-than-Nyquist signaling to improve spectral efficiency. To our knowledge, MOCZ has not been studied for OAC in the literature. 

{\color{\reviewColor}
 In this work, inspired by MOCZ, we  strive to answer how the zeros of polynomials can be utilized for OAC. Using the zeros of polynomials for OAC is challenging since the signal superposition changes the original zeros of the polynomials at the transmitters non-linearly. 
   %Although the existing literature on MOCZ discusses the communication aspects in detail, it is not trivial to extend them for OAC. Likewise, the existing literature on OAC discusses MV computation, they still do not provide insight into this challenge. 
 Our main contribution lies in addressing this mathematical challenge and showing that we can encode the zeros of polynomials for OAC without using the CSI at the transmitters or receiver.   Our specific contributions can be listed as follows:}
\begin{itemize}
	\item We introduce a new approach to compute the \ac{MV} function based on \ac{MOCZ}. The key property that we use is that when a linear combination of polynomials is evaluated at a specific value, and if this particular  value corresponds to a zero of a polynomial in the combinations, the contribution of that polynomial to the evaluation is zero. 
	{\color{\reviewColor}Based on this property % the transmitter chooses the zeros of Huffman's polynomials based on its votes, i.e., $+1$ and $-1$, and the receiver evaluates the polynomial constructed with the elements of the superposed sequence at the corresponding zeros.
	and Lemmas~{\ref{lemma:exp}-\ref{lemma:exp3}}, we prove that a non-coherent aggregation can be achieved in the fading channel and a  \ac{dizet} decoder, initially proposed for communications with MOCZ in \cite{Philipp_2019principles}, can be used to obtain the MVs.}

	\item We propose three methods, each with its own advantages. While Method~1 provides the highest computation rate, it requires delay profile information. We address this issue by using a differential encoding strategy in Method~2 at the expense of halved computation rate. Finally, by extending our preliminary results in \cite{sahinGC_2024}, we reduce the \ac{CER} by introducing redundancy in Method~3 with an index-based encoder. {\color{\reviewColor}Also, we  analytically derive the \acp{CER} in Corollaries~\ref{corr:cermethod1}-\ref{corr:cerMethod3} based on Lemma~\ref{lemma:errProb}.} All methods are robust to time and phase synchronization errors as they do not rely on the availability of \ac{CSI} at the transmitters and receiver.

	\item {\color{\reviewColor}We assess each method's performance with comprehensive simulations and demonstrate the applicability of the proposed methods to a distributed median computation scenario using MVs.}
	%We also generate numerical results based on Goldenbaum's \ac{OAC} scheme in \cite{Goldenbaum_2013tcom}.
	
\end{itemize}

{\em Organization:} The rest of the paper is organized as follows. Section~\ref{sec:system} provides the system model. 
In Section~\ref{sec:scheme}, we discuss the proposed \ac{OAC} methods in detail. In Section~\ref{sec:performance}, we theoretically analyze the \acp{CER} of the proposed methods.
 In Section~\ref{sec:numerical}, we provide numerical results and assess the methods in a distributed median computation scenario. We conclude the paper in Section~\ref{sec:conclusion}.

{\em Notation:} The sets of complex and real numbers are denoted by $\complexNumbers$ and  $\realNumbers$, respectively.
The function $\signNormal[\cdot]$ results in $1$, $-1$, or $0$ for a positive, a negative, or a zero-valued argument, respectively. 
%%The $N$-dimensional all zero vector and the $N\times N$ identity matrix are  $\zeroVector[{N}]$ and $\identityMatrix[{N}]$, respectively. 
%%The function $\indicatorFunction[\cdot]$ results in $1$ if its argument holds, otherwise, it is $0$. 
$\expectationOperator[\cdot][x]$ is the expectation of its  argument over all random variables. 
%%The operation $\diagOperator[\textbf{a}]$  returns a square diagonal matrix with the elements of vector \textbf{a} on the main diagonal.
%%$ \nabla \lossFunctionSample[{\modelParameters}]$ denotes the gradient of the function $f$, i.e. $\nabla f$, at the point $\modelParameters$. 
The zero-mean circularly symmetric complex Gaussian distribution with variance $\sigma^2$ is denoted by $\complexGaussian[0][\sigma^2]$. 
%%$\vectorX\sim\complexGaussian[\zeroVector[\numberOfActiveSubcarriers]][{\textbf{\textrm{C}}_{\numberOfActiveSubcarriers}}]$. The gamma distribution with the shape parameter $n$ and the rate  $\lambda$ is  $\gammaDist[n][\lambda]$.
%%The binomial distribution with the $K$ trials and the success probability  $p$ for each trial is $\binomDist[K][p]$.
The uniform distribution with the support between $a$ and $b$ is $\uniformDistribution[a][b]$. 
%%Normal distribution with mean $\mu$ and variance $\sigma^2$ is $\gaussianDist[\mu][\sigma^2]$.
%%The $\ell_2$-norm of the vector $\vectorX$ is $\norm{\vectorX}_2$. 
%The cumulative distribution function of the standard normal distribution is $\normalCDF[\cdot]$. 
The function $\indicatorFunction[\cdot]$ results in $1$ if its argument holds; otherwise, it is $0$.
The probability of  an event $A$ is denoted by $\probability[A;x]$, where $x$ is a parameter to calculate the probability. 
%The conditional probability of an event $A$ given the event $B$ is shown as $\probability[A|B]$. $\oneVector[L]$ and $\zeroVector[L]$ denote the vectors of length $L$, where their elements are only $1$ or $0$, respectively.

\section{System Model}
\label{sec:system}
Consider an \ac{OAC} scenario with $\numberOfEdgeDevices$ transmitters and a receiver, where all radios are equipped with a single antenna.  {\color{\reviewColor}Let  $\transmittedSeqEle[\indexED][\indexTime]\in\complexNumbers$ be the $\indexTime$th element of sequence $\transmittedSeq[\indexED]$ to be transmitted from the $\indexED$th transmitter over an orthogonal waveform. 
Assume that the impact of the composite  channel between the $\indexED$th transmitter and the receiver on the received sequence  is convolutive and the corresponding impulse response is  $\channelSeq[\indexED]=(\channelSeqEle[\indexED][0],\mydots,\channelSeqEle[\indexED][\channelLength-1])$, where $\channelLength\ge1$ is the number of effective taps.\footnote{\color{\reviewColor} The composite channel is a function of the multipath channel and the waveform. For instance, for OFDM,  each subcarrier is exposed to a single-tap channel (i.e., $\channelLength=1$) regardless of the \acl{PDP} of the multipath channel if the cyclic prefix duration is large enough.  On the other hand,  $\channelLength$ can be larger than $1$ for a single-carrier waveform after matched filtering.} All transmitters access the medium concurrently for computation.} We can then express the $\indexTime$th received element $\receivedSeqEle[\indexTime]\in\complexNumbers$ at the receiver after the signal superposition as
\begin{align}
	\receivedSeqEle[\indexTime]
	=\left(\sum_{\indexED=1}^{\numberOfEdgeDevices}
	\sum_{\indexTap=0}^{\channelLength-1}
		\sqrt{\transmitPower[\indexED]}\channelSeqEle[\indexED][\indexTap] 
		\transmittedSeqEle[\indexED][\indexTime-\indexTap]\right)+\noiseSample[\indexTime]~,	\label{eq:receivedSamples}
\end{align}
where $\channelSeqEle[\indexED][\indexTap]\sim\complexGaussian[0][{\pdpTap[\indexTap]}]$ is the channel coefficient on the $\indexTap$th tap for the $\indexED$th transmitter, $\transmitPower[\indexED]$ is the average transmit power of the $\indexED$th transmitter, and ${\noiseSample[\indexTime]}\sim\complexGaussian[0][\noiseVariance]$ is the \ac{AWGN}. {\color{\reviewColor}Without loss of generality,} we consider exponential decay to model the composite channel between the $\indexED$th transmitter and the receiver as
\begin{align}
	\pdpTap[\indexTap] = \expectationOperator[{|\channelSeqEle[\indexED][\indexTap]|^2}][]=
	\begin{cases}
	\frac{1-\decayingParameter}{1-\decayingParameter^{\channelLength}}\decayingParameter^\indexTap~,&\decayingParameter<1\\
	\frac{1}{\channelLength}~,&\decayingParameter=1
	\end{cases}~,
\end{align}
 for $\sum_{\indexTap=0}^{\channelLength-1}\pdpTap[\indexTap]=1$, where $\decayingParameter\in(0,1]$ is a decay constant  \cite{Philipp_2019principles}.

 We assume that the {\em average} received signal powers of the transmitters are aligned with a power control mechanism \cite{10.5555/3294673}. Thus, the relative positions of the transmitters to the receiver do not change our analyses as in \cite{Guangxu_2020, Guangxu_2021, Busra_2023}. Also,  we set $\transmitPower[\indexED]$, $\forall\indexED$, to $1$~Watt and calculate the average \ac{SNR} of a transmitter at the receiver as $\SNR=1/\noiseVariance$.

Let $\codedSeq[\indexED]=(\codedSeqEle[\indexED][0],\codedSeqEle[\indexED][1],\dots, \codedSeqEle[\indexED][\numberOfRoots])$  denote the complex-valued coefficients of the polynomial function 
$\polySeq[\codedSeqP[\indexED]][\polyVariable] = \codedSeqEle[\indexED][\numberOfRoots]\polyVariable^{\numberOfRoots} + \codedSeqEle[\indexED][\numberOfRoots-1]\polyVariable^{\numberOfRoots-1}+ \dots + \codedSeqEle[\indexED][0]$
for $\polyVariable\in\complexNumbers$ and $\codedSeqEle[\indexED][\numberOfRoots]\neq0$.
%\footnote{In the literature, $\polyVariable$-transform is defined with the negative exponent. In this study, we use positive exponent for simplifying the notation.} 
We set $\transmittedSeqEle[\indexED][\indexTime]$ as 
\begin{align}
	\transmittedSeqEle[\indexED][\indexTime] = 
	\begin{cases}
		\codedSeqEle[\indexED][\indexTime],& 0\le\indexTime \le\numberOfRoots \\
		0,& \text{otherwise}
	\end{cases}.
\end{align}
Since a convolution operation in discrete time can be represented by a polynomial multiplication in the $\polyVariable$-domain, we can express the  received sequence $\receivedSeq[]=(\receivedSeqEle[0],\mydots,\receivedSeqEle[\numberOfRoots+\channelLength-1])$ in \eqref{eq:receivedSamples}  as
\begin{align}
	\polySeq[{\receivedSeqP}][\polyVariable]
	&=\underbrace{\sum_{\indexED=1}^{\numberOfEdgeDevices}
	\polySeq[\codedSeqP[\indexED]][\polyVariable]\polySeq[\channelSeqP[\indexED]][\polyVariable]}_{\triangleq\polySeq[\superposedSeqP][\polyVariable]}+\polySeq[\noiseSeqP][\polyVariable]~,
	\label{eq:rxPoly}
\end{align}
where $\polySeq[\noiseSeqP][\polyVariable]$ is the $\polyVariable$-domain representation of the noise sequence $\noiseSeq=(\noiseSeqEle[0],\mydots,\noiseSeqEle[\numberOfRoots+\channelLength-1])$, and  $\polySeq[\codedSeqP[\indexED]][\polyVariable]$ and $\polySeq[\channelSeqP[\indexED]][\polyVariable]$ are the $\polyVariable$-domain representations of $\codedSeq[\indexED]$ and $\channelSeq[\indexED]$, respectively, i.e.,
\begin{align}
	\polySeq[\codedSeqP[\indexED]][\polyVariable]=\sum_{\indexTime=0}^{\numberOfRoots}\codedSeqEle[\indexED][\indexTime]\polyVariable^{\indexTime}=
	\codedSeqEle[\indexED][\numberOfRoots]\prod_{\indexRoot=0}^{\numberOfRoots-1} (\polyVariable-\rootCodedSeq[\indexED][\indexRoot])
	~,
	\label{eq:polyDef}
\end{align}
and
\begin{align}
	\polySeq[\channelSeqP[\indexED]][\polyVariable]=\sum_{\indexTap=0}^{\channelLength-1}\channelSeqEle[\indexED][\indexTap]\polyVariable^{\indexTap}=
	\channelSeqEle[\indexED][\channelLength-1]\prod_{\indexTap=0}^{\channelLength-2} (\polyVariable-\rootChannelSeq[\indexED][\indexTap])~,
\end{align}
by using the facts that $\polySeq[\codedSeqP[\indexED]][\polyVariable]$ and $\polySeq[\channelSeqP[\indexED]][\polyVariable]$ have $\numberOfRoots$ and $\channelLength-1$ complex-valued roots, respectively, by the fundamental theorem of algebra.

{\color{\reviewColor}
	\subsection{Preliminaries on Huffman Sequences}
Let $\seqGx\triangleq(\eleGx[0],\eleGx[1],\dots, \eleGx[\numberOfRoots])$ denote the complex-valued coefficients of the polynomial function 
$\polySeq[\seqGxP][\polyVariable] =\sum_{\indexTime=0}^{\numberOfRoots}\eleGx[\indexTime]\polyVariable^{\indexTime}   =	\eleGx[\numberOfRoots]\prod_{\indexRoot=0}^{\numberOfRoots-1} (\polyVariable-\rootSeqX[\indexRoot])$,  for $\eleGx[\numberOfRoots]\neq0$. 
We define the \ac{AACF} of the sequence $\seqGx$ as $	\polySeq[\seqGaP][\polyVariable] \triangleq \polyVariable^{\numberOfRoots}\sum_{\lagForCorrelation=-\numberOfRoots}^{\numberOfRoots} \apac[\seqGx][\lagForCorrelation] \polyVariable^{\ell}
$,
where $\apac[\seqGx][\lagForCorrelation]$ is the $\lagForCorrelation$th aperiodic auto-correlation coefficient of the sequence $\seqGx$, given by
\begin{align}
	\apac[\seqGx][\lagForCorrelation]\triangleq
	\begin{cases}
		\sum_{\indexEleOfSeq=0}^{\numberOfRoots-\lagForCorrelation} \eleGx[\indexEleOfSeq]^*\eleGx[\indexEleOfSeq+\lagForCorrelation], & 0\le\lagForCorrelation\le\numberOfRoots\\
		\sum_{\indexEleOfSeq=0}^{\numberOfRoots+\lagForCorrelation} \eleGx[\indexEleOfSeq]\eleGx[\indexEleOfSeq-\lagForCorrelation]^*, & -\numberOfRoots\le\lagForCorrelation<0\\
		0,& \text{otherwise}
	\end{cases}~.
\end{align}
A Huffman sequence and a Huffman polynomial can then be defined as follows:
\begin{definition}[Huffman sequence and Huffman polynomial]
	 The sequence $\seqGx$ and the polynomial  $\polySeq[\seqGxP][\polyVariable]$ are called a Huffman sequence and Huffman polynomial, respectively, if $\polySeq[\seqGaP][\polyVariable]=\bar{\lastCorrTerm}+\norm{\seqGx}_2^2\polyVariable^{\numberOfRoots}+\lastCorrTerm\polyVariable^{2\numberOfRoots}$ for $\lastCorrTerm=\eleGx[\numberOfRoots]\eleGxConj[0]$.
\end{definition}

Notice that the \ac{AACF} $\polySeq[\seqGaP][\polyVariable]$ can be directly calculated from the zeros of $\polySeq[\seqGxP][\polyVariable]$ as  \cite{Philipp_2019principles}
\begin{align}
	\polySeq[\seqGaP][\polyVariable]  =& \polyVariable^{\numberOfRoots}\polySeq[\seqGxP][\polyVariable]\polySeqConj[\seqGxP][1/\bar{\polyVariable}]\nonumber\\=&\eleGx[\numberOfRoots]\eleGxConj[0]\prod_{\indexRoot=0}^{\numberOfRoots-1} (\polyVariable-\rootSeqX[\indexRoot])\prod_{\indexRoot=0}^{\numberOfRoots-1} (\polyVariable-\frac{1}{\rootSeqXConj[\indexRoot]})~.
	\label{eq:apac}
\end{align}
In \cite{Huffman_1962}, by exploiting the $\numberOfRoots$th root of unity and conjugate-reciprocal zero pairs of $\polySeq[\seqGaP][\polyVariable]$ in \eqref{eq:apac}, i.e., $\{(\rootSeqX[\indexRoot],\frac{1}{\rootSeqXConj[\indexRoot]}),\forall\indexRoot\}$, Huffman shows that 
 the $\numberOfRoots$ zeros of a Huffman polynomial are evenly placed on two reciprocal circles centered at the origin, 
 %where the angles of zeros uniformly divide the $0$ to $2\pi$ range into $\numberOfRoots$ angles, 
 and their amplitudes  can be either $\radiusRoot$ or $\radiusRoot^{-1}$ for $\radiusRoot>1$, i.e.,  $\rootSeqX[\indexRoot]\in\setHuffmanRoots[\indexRoot]\triangleq\{\radiusRoot\constante^{\frac{\constantj2\pi\indexRoot}{\numberOfRoots}},\radiusRoot^{-1}\constante^{\frac{\constantj2\pi\indexRoot}{\numberOfRoots}}\}$. Thus, for a given $\radiusRoot$, $2^\numberOfRoots$ distinct Huffman sequences with an identical \ac{AACF} can be synthesized. 
 
 The \acp{AACF} of Huffman sequences are  very close to impulse function. Hence, they can be useful for radar applications \cite{Martin_1970}. In \cite{Philipp_2019principles} and \cite{Philipp_2020practical}, it is shown that the zeros of Huffman polynomials are numerically stable when the corresponding polynomial coefficients are perturbed with additive noise and can be utilized for modulation, leading to \ac{BMOCZ}. At the transmitter, an information bit is encoded into one of the zeros in a conjugate-reciprocal zero pair. At the receiver side, a low-complexity non-coherent detector that compares two metrics by evaluating the polynomial at the zeros of a conjugate-reciprocal zero pair, i.e., \ac{dizet} detector, is employed to detect the information bits without \ac{CSI}. 
% by showing that Huffman polynomials provide robust against additive noise.

Given their numerical stability, we consider Huffman sequences for  $\codedSeq[\indexED]$ in this work. We normalize  $\norm{\codedSeq[\indexED]}_2^2$  to $\numberOfRoots+1$ by setting $\codedSeqEle[\indexED][\numberOfRoots]$ (see \eqref{eq:polyDef}) as
\begin{align}
	\codedSeqEle[\indexED][\numberOfRoots]=\sqrt{\frac{\lastCorrTerm({\numberOfRoots+1})}{\prod_{\indexRoot=0}^{\numberOfRoots-1}|\rootCodedSeq[\indexED][\indexRoot]|}}~,
	\label{eq:normalization}
\end{align}
for $\lastCorrTerm={1}/({\radiusRoot^\numberOfRoots+\radiusRoot^{-\numberOfRoots}})$ and  $\radiusRoot\triangleq\sqrt{1+\sin(\pi/\numberOfRoots)}$. Note that this specific value of $\radiusRoot$ maximizes the minimum distance between the zeros to improve the robustness against noise for a \ac{dizet} decoder \cite{Philipp_2019principles}.}

\subsection{Problem Statement}
Suppose that the fading coefficients, i.e., \{$\channelSeq[\indexED]$, $\forall\indexED$\}, are not available at the transmitters and the receiver, and the receiver is interested in computing $\numberOfMVs$ \ac{MV} functions expressed as
\begin{align}
\majorityVoteEle[\indexMV]=\signNormal[{\sum_{\indexED=1}^{\numberOfEdgeDevices}{\voteVectorEDEle[\indexED][\indexMV]}}]~,\quad \forall\indexMV\in\{0,1,\mydots,\numberOfMVs-1\},
\label{eq:mvProblem}
\end{align}
where $\voteVectorEDEle[\indexED][\indexMV]\in\{-1,1\}$ represents the $\indexMV$th vote of $\indexED$th transmitter and $\majorityVoteEle[\indexMV]\in\{-1,1\}$ is the $\indexMV$th \ac{MV}.  {\em \color{\reviewColor} How can the \acp{MV}  be calculated with OAC by exploiting the zeros of polynomials while still being agnostic to the \ac{CSI} at the transmitters and receiver?} Although \ac{MOCZ} allows the receiver to use non-coherent detectors to obtain the bits for single-user {\em communications}, it is not trivial to use the same concept for {\em computation} in the channel as the  signal superposition for $\numberOfEdgeDevices>1$  in \eqref{eq:rxPoly} changes the original zeros of the polynomials at the transmitters non-linearly.
{\color{\reviewColor}\begin{example}
	Consider the Huffman polynomials given by
	\begin{align}
	\polySeq[\codedSeqP[1]][\polyVariable]=\sqrt{\frac{12}{17}}(\polyVariable-\frac{1}{2})(\polyVariable+2) = \sqrt{\frac{12}{17}}(\polyVariable^2+\frac{3}{2}\polyVariable-1)~,\nonumber\\
		\polySeq[\codedSeqP[1]][\polyVariable]=\sqrt{\frac{12}{17}}(\polyVariable-2)(\polyVariable+\frac{1}{2}) = \sqrt{\frac{12}{17}}(\polyVariable^2-\frac{3}{2}\polyVariable-1)~,\nonumber
	\end{align}
for $\lastCorrTerm=4/17$, $\radiusRoot=2$, and $\numberOfRoots=2$. In an ideal channel, the sum of these polynomials can be obtained as
\begin{align}
\polySeq[\superposedSeqP][\polyVariable]&=\polySeq[\codedSeqP[1]][\polyVariable]+\polySeq[\codedSeqP[2]][\polyVariable] =2\sqrt{\frac{12}{17}}(\polyVariable-1)(\polyVariable+1)\nonumber~,
\end{align}
which demonstrates that the zeros of $\polySeq[\superposedSeqP][\polyVariable]$ are non-trivial functions of the zeros of $\polySeq[\codedSeqP[1]][\polyVariable]$ and $\polySeq[\codedSeqP[2]][\polyVariable]$.
\end{example}}

\section{Proposed Methods}
\label{sec:scheme}In this section, we discuss three methods to compute MVs.
{\color{\reviewColor} We introduce Lemmas 1-3, which allow us to derive the statistics that estimate the number of nodes voting for $1$ and $-1$ and the corresponding  detectors.} For the derivations, we need the following functions related to channel and noise:
\begin{align}
		\funcChannel[\radiusRoot]&\triangleq\expectationOperator[|{\polySeq[\channelSeqP[\indexED]][{
			\radiusRoot\constante^{\frac{\constantj2\pi\indexMV}{\numberOfRoots}}
		}]}|^2][]=\sum_{\indexTap=0}^{\channelLength-1}\expectationOperator[{|\channelSeqEle[\indexED][\indexTap]
		\radiusRoot^{\indexTap}\constante^{\frac{\constantj2\pi\indexMV\indexTap}{\numberOfRoots}}
		|^2}][]\nonumber\\&=\sum_{\indexTap=0}^{\channelLength-1}\pdpTap[\indexTap] \radiusRoot^{2\indexTap}
=
	\begin{cases}
		\frac{1-\decayingParameter}{1-\decayingParameter^{\channelLength}}\frac{1-\radiusRoot^{2\channelLength}\decayingParameter^{\channelLength}}{1-\radiusRoot^2\decayingParameter}~,&\decayingParameter<1\\
		\frac{1}{\channelLength}\frac{1-\radiusRoot^{2\channelLength}}{1-\radiusRoot^2}~,&\decayingParameter=1
	\end{cases}~.
	\label{eq:expectedChannel}
\end{align}
and
\begin{align}
	\funcNoise[\radiusRoot] \triangleq	\expectationOperator[\left|{\polySeq[\noiseSeqP][ \radiusRoot\constante^{\frac{\constantj2\pi\indexMV}{\numberOfRoots}}]}\right|^2 ][] 
	=\noiseVariance\frac{1-\radiusRoot^{2(\numberOfRoots+\channelLength)}}{1-\radiusRoot^2}~.
\end{align}
{\color{\reviewColor}It is worth noting that an arbitrary delay profile can also be considered for $\funcChannel[\radiusRoot]$ without changing the derivations in the following subsections.

}

\subsection{Method 1: Uncoded MV Computation}
In this method,  we compute $\numberOfMVs=\numberOfRoots$ \acp{MV} without any coding and the $\indexED$th transmitter sets the $\indexRoot$th root of $\polySeq[\codedSeqP[\indexED]][\polyVariable]$ based on the vote $\voteVectorEDEle[\indexED][\indexMV]$, $\forall\indexMV\in\{0,1,\mydots,\numberOfRoots-1\}$, as
\begin{align}
\rootCodedSeq[\indexED][\indexRoot]
&=\begin{cases}
		\frac{1}{\radiusRoot}\constante^{\frac{\constantj2\pi\indexRoot}{\numberOfRoots}}\quad,&\voteVectorEDEle[\indexED][\indexRoot]=1\\
\radiusRoot\constante^{\frac{\constantj2\pi\indexRoot}{\numberOfRoots}}\quad,&\voteVectorEDEle[\indexED][\indexRoot]=-1\\
\end{cases}~.
\label{eq:schemeOne}
\end{align}
The encoding in \eqref{eq:schemeOne} is very similar to \ac{BMOCZ} in \cite{Philipp_2019principles}. However, since the transmitted signals superpose for OAC in \eqref{eq:rxPoly}, it is not trivial how to design the detector to detect the MVs. We use the following lemma to develop the decoder:
\begin{lemma}
Let $\numberOfEdgeDevicesPlus[\indexMV]$ and $\numberOfEdgeDevicesMinus[\indexMV]$ denote the number of transmitters with positive and negative votes for the $\indexMV$th MV computation.  For the mapping in \eqref{eq:schemeOne} with $\probability[{\voteVectorEDEle[\indexED][\indexMV']=1}]=\probability[{\voteVectorEDEle[\indexED][\indexMV']=-1}]=1/2$, $\forall\indexMV'$, $\indexMV'\neq\indexMV$,
	\begin{align}
		\expectationOperator[\left|{\polySeq[\superposedSeqP][ \radiusRoot\constante^{\frac{\constantj2\pi\indexMV}{\numberOfRoots}}]}\right|^2 ][] &= \numberOfEdgeDevicesPlus[\indexMV]
		\funcSignalOne[\radiusRoot]
		\funcChannel[\radiusRoot]~,
		\label{eq:expectedPlus}
	\end{align}
	where
	\begin{align}
	&\funcSignalOne[\radiusRoot]\triangleq\lastCorrTerm({\numberOfRoots+1})(\radiusRoot-\radiusRoot^{-1})^2\radiusRoot^{{\numberOfRoots}}\nonumber
	\\&~~~~~~~
	\times\frac{1}{2^{\numberOfRoots-1}}\prod_{\substack{\indexRoot=1}}^{\numberOfRoots-1} {{|1-\constante^{\frac{\constantj2\pi\indexRoot}{\numberOfRoots}} |^2} + {|\radiusRoot-\radiusRoot^{-1}\constante^{\frac{\constantj2\pi\indexRoot}{\numberOfRoots}} |^2}}~,
\end{align}	
where the expectation in \eqref{eq:expectedPlus} is over the distributions of channels and votes.
	\label{lemma:exp}
\end{lemma}
The proof is given in Appendix \ref{app:lemma1}.
\begin{corollary}
 $\expectationOperator[\left|{\polySeq[\superposedSeqP][ \radiusRoot^{-1}\constante^{\frac{\constantj2\pi\indexMV}{\numberOfRoots}}]}\right|^2 ][]$ can be calculated by replacing  with  $\numberOfEdgeDevicesPlus[\indexMV]$ and $\radiusRoot$ with $\numberOfEdgeDevicesMinus[\indexMV]$ and $\radiusRoot^{-1}$ in \eqref{eq:expectedPlus}, respectively.
 \label{corr:exp}
\end{corollary}

The key observation from Lemma~\ref{lemma:exp} and Corollary~\ref{corr:exp} is that the expected values of $|{\polySeq[\superposedSeqP][ \radiusRoot\constante^{\frac{\constantj2\pi\indexMV}{\numberOfRoots}}]}|^2$ and $|{\polySeq[\superposedSeqP][ \radiusRoot^{-1}\constante^{\frac{\constantj2\pi\indexMV}{\numberOfRoots}}]}|^2$ are linearly scaled by $\numberOfEdgeDevicesPlus[\indexMV]$ and $\numberOfEdgeDevicesMinus[\indexMV]$, respectively. Thus, a \ac{dizet} decoder, initially  used for detecting bits \cite{Philipp_2019principles}, can still be utilized to compute the $\indexMV$th \ac{MV} with proper scalars highlighted by Lemma~\ref{lemma:exp} and Corollary~\ref{corr:exp} as
\begin{align}
	\majorityVoteDetectedEle[\indexMV]=\sign{(\metricPlus[\indexMV]-\metricMinus[\indexMV])}~,
	\label{eq:detector1}
\end{align}
where $\metricPlus[\indexMV]$ and $\metricMinus[\indexMV]$ are the unbiased estimates of $\numberOfEdgeDevicesPlus[\indexMV]$ and $\numberOfEdgeDevicesMinus[\indexMV]$, respectively, given by
\begin{align}
	\metricPlus[\indexMV] =		\frac{\left|{\polySeq[\receivedSeqP][ \radiusRoot\constante^{\frac{\constantj2\pi\indexMV}{\numberOfRoots}}]}\right|^2-\funcNoise[\radiusRoot]}{\funcSignalOne[\radiusRoot]
		\funcChannel[\radiusRoot]}~,
		\label{eq:upOne}
\end{align}
and
\begin{align}
	\metricMinus[\indexMV]=
\frac{\left|{\polySeq[\receivedSeqP][ \radiusRoot^{-1}\constante^{\frac{\constantj2\pi\indexMV}{\numberOfRoots}}]}\right|^2-\funcNoise[\radiusRoot^{-1}]}{\funcSignalOne[\radiusRoot^{-1}]
	\funcChannel[\radiusRoot^{-1}]}~.
	\label{eq:umOne}
\end{align}
Note that \eqref{eq:detector1} can be simplified by using  ${\funcSignalOne[\radiusRoot]}/{\funcSignalOne[\radiusRoot^{-1}]
}=\radiusRoot^{2\numberOfRoots}$, but it still requires the delay profile of the channel. 
{\color{\reviewColor}\begin{example}
	For $\radiusRoot=2$ and $\numberOfRoots=2$, $\funcSignalOne[\radiusRoot]$ and $\funcSignalOne[\radiusRoot^{-1}]$ are approximately $32.5588$ and $2.0349$, respectively, and their ratio is $16$.
\end{example}}

The computation rate for Method~1 can also obtained as $\numberOfRoots$ \acp{MV} over $\numberOfRoots+\channelLength$ complex-valued resources. The number of consumed resources is $\numberOfRoots+\channelLength$ as the transmitted sequence needs to be padded with $\channelLength$ zeros to express \eqref{eq:rxPoly}.

\subsection{Method 2: Differential MV Computation}
In this method, we use differential encoding to compute $\numberOfMVs=\numberOfRoots/2$ \acp{MV} and the $\indexED$th transmitter sets the $\indexRoot$th and $(\indexRoot+1)$th roots of $\polySeq[\codedSeqP[\indexED]][\polyVariable]$ based on the vote $\voteVectorEDEle[\indexED][\indexMV]$, $\forall\indexMV\in\{0,1,\mydots,\numberOfRoots/2-1\}$, as
\begin{align}
(\rootCodedSeq[\indexED][2\indexRoot'],\rootCodedSeq[\indexED][2\indexRoot'+1])
&=\begin{cases}
(	\frac{1}{\radiusRoot}\constante^{\frac{\constantj2\pi2\indexRoot'}{\numberOfRoots}},\radiusRoot\constante^{\frac{\constantj2\pi(2\indexRoot'+1)}{\numberOfRoots}}),&\voteVectorEDEle[\indexED][\indexRoot']=1\\
	(\radiusRoot\constante^{\frac{\constantj2\pi2\indexRoot'}{\numberOfRoots}},\frac{1}{\radiusRoot}\constante^{\frac{\constantj2\pi(2\indexRoot'+1)}{\numberOfRoots}}),&\voteVectorEDEle[\indexED][\indexRoot']=-1\\
\end{cases}~.
\label{eq:schemeTwo}
\end{align}
for $\indexRoot' = \floor{{\indexRoot}/{2}}$.
%\begin{align}
%	\rootCodedSeq[\indexED][\indexRoot]
%	&=\begin{cases}
%	\frac{1}{\radiusRoot}\constante^{\frac{\constantj2\pi\indexRoot}{\numberOfRoots}},&\voteVectorEDEle[\indexED][\floor{\frac{\indexRoot}{2}}]=1, \indexRoot \text{ is even}\\
%	\radiusRoot\constante^{\frac{\constantj2\pi\indexRoot}{\numberOfRoots}},&\voteVectorEDEle[\indexED][\floor{\frac{\indexRoot}{2}}]=1, \indexRoot \text{ is odd}\\
%		\radiusRoot\constante^{\frac{\constantj2\pi\indexRoot}{\numberOfRoots}},&\voteVectorEDEle[\indexED][\floor{\frac{\indexRoot}{2}}]=-1, \indexRoot \text{ is even}\\
%		\frac{1}{\radiusRoot}\constante^{\frac{\constantj2\pi\indexRoot}{\numberOfRoots}},&\voteVectorEDEle[\indexED][\floor{\frac{\indexRoot}{2}}]=-1, \indexRoot \text{ is odd}\\
%	\end{cases}~.
%	\label{eq:schemeTwo}
%\end{align} 
To derive the corresponding detector, we use the following lemma:
\begin{lemma}
Let $\numberOfEdgeDevicesPlus[\indexMV]$ and $\numberOfEdgeDevicesMinus[\indexMV]$ denote the number of transmitters with positive and negative votes for $\indexMV$th MV computation.  For the mapping in \eqref{eq:schemeTwo} with $\probability[{\voteVectorEDEle[\indexED][\indexMV']=1}]=\probability[{\voteVectorEDEle[\indexED][\indexMV']=-1}]=1/2$, $\forall\indexMV'$, $\indexMV'\neq\indexMV$,	
	\begin{align}
		\expectationOperator[\left|{\polySeq[\superposedSeqP][ \radiusRoot\constante^{\frac{\constantj2\pi2\indexMV}{\numberOfRoots}}]}\right|^2 ][] &= \numberOfEdgeDevicesPlus[\indexMV]
		\funcSignalTwo[\radiusRoot]
		\funcChannel[\radiusRoot]~,
		\label{eq:expectedPlusP1}
	\end{align}
	where
	\begin{align}
		&\funcSignalTwo[\radiusRoot]\triangleq	\lastCorrTerm({\numberOfRoots+1})(\radiusRoot-\radiusRoot^{-1})^2|1-\constante^{\frac{2\pi}{\numberOfRoots}}|^2\nonumber\radiusRoot^{\numberOfRoots}
		\\&~~~~~~
\times\frac{1}{2^{\frac{\numberOfRoots}{2}-1}}\prod_{\substack{\indexRoot=1}}^{\frac{\numberOfRoots}{2}-1} {{|1-\constante^{\frac{\constantj2\pi2\indexRoot}{\numberOfRoots}} |^2}} {{|\radiusRoot-\radiusRoot^{-1}\constante^{\frac{\constantj2\pi(2\indexRoot+1)}{\numberOfRoots}} |^2}}
\nonumber	\\
&~~~~~~~~~~~~~~~~~~~~+ 
{|1-\constante^{\frac{\constantj2\pi(2\indexRoot+1)}{\numberOfRoots}} |^2}
{{|\radiusRoot-\radiusRoot^{-1}\constante^{\frac{\constantj2\pi2\indexRoot}{\numberOfRoots}} |^2}}~, 
	\end{align}	
	where the expectation in \eqref{eq:expectedPlusP1}  is over the distributions of channels and votes.
	\label{lemma:exp2}
\end{lemma}
The proof is given in Appendix \ref{app:lemma2}.

\begin{corollary}
 $\expectationOperator[\left|{\polySeq[\superposedSeqP][ \radiusRoot\constante^{\frac{\constantj2\pi(2\indexMV+1)}{\numberOfRoots}}]}\right|^2 ][]$ can be calculated by replacing   $\numberOfEdgeDevicesPlus[\indexMV]$ with $\numberOfEdgeDevicesMinus[\indexMV]$ in \eqref{eq:expectedPlusP1}.
	\label{corr:exp2}
\end{corollary}

With Lemma~\ref{lemma:exp2} and Corollary~\ref{corr:exp2}, the unbiased estimates of $\numberOfEdgeDevicesPlus[\indexMV]$ and $\numberOfEdgeDevicesMinus[\indexMV]$ can be obtained as
\begin{align}
	\metricPlus[\indexMV] =	
	\frac{\left|{\polySeq[\receivedSeqP][ \radiusRoot\constante^{\frac{\constantj2\pi2\indexMV}{\numberOfRoots}}]}\right|^2
	-\funcNoise[\radiusRoot]}{
	\funcSignalTwo[\radiusRoot]
	\funcChannel[\radiusRoot] 
}
~,	
\end{align}
and
\begin{align}
	\metricMinus[\indexMV]=	
		\frac{\left|{\polySeq[\receivedSeqP][ \radiusRoot\constante^{\frac{\constantj2\pi(2\indexMV+1)}{\numberOfRoots}}]}\right|^2-\funcNoise[\radiusRoot]}{
	\funcSignalTwo[\radiusRoot]
	\funcChannel[\radiusRoot] 
}
~,
\end{align}
respectively. Hence, the $\indexMV$th \ac{MV} can be computed as 
\begin{align}
	\majorityVoteDetectedEle[\indexMV]=&\text{sign}\left(\left|{\polySeq[\receivedSeqP][ \radiusRoot\constante^{\frac{\constantj2\pi(2\indexMV+1)}{\numberOfRoots}}]}\right|^2 
				-\left|{\polySeq[\receivedSeqP][ \radiusRoot\constante^{\frac{\constantj2\pi2\indexMV}{\numberOfRoots}}]}\right|^2
			\right).
			\label{eq:detector2}
\end{align}
Compared with the detector in Method 1,  the detector in \eqref{eq:detector2} does not need the delay profile of the channel to compute the \acp{MV}. The price paid for this benefit is a reduced computation rate, i.e., $\numberOfRoots/2$ \acp{MV} over $\numberOfRoots+\channelLength$ complex-valued resources.

\subsection{Method 3: Index-based MV Computation}
In this approach, we compute $\numberOfMVs=\log_2(\numberOfRoots)$ \acp{MV}, and the  roots of $\polySeq[\codedSeqP[\indexED]][\polyVariable]$ are modulated based on an index calculated by using all votes. To this end, let
$\voteVectorEDBinaryEle[\indexED][\indexMV]\in\integers_2$ be a binary representation of the vote $\voteVectorEDEle[\indexED][\indexMV]$ as $\voteVectorEDBinaryEle[\indexED][\indexMV]\triangleq(\voteVectorEDEle[\indexED][\indexMV]+1)/2$ for $\forall\indexMV\in\{0,1,\mydots,\log_2(\numberOfRoots)-1\}$. The $\indexED$th transmitter sets the $\indexRoot$th root of $\polySeq[\codedSeqP[\indexED]][\polyVariable]$ as
\begin{align}
	\rootCodedSeq[\indexED][\indexRoot]
	&=\begin{cases}			  \frac{1}{\radiusRoot}\constante^{\frac{\constantj2\pi\indexRoot}{\numberOfRoots}},&\sum_{\indexMV=0}^{\numberOfMVs-1}\voteVectorEDBinaryEle[\indexED][\indexMV]2^\indexMV=\indexRoot\\
		\radiusRoot\constante^{\frac{\constantj2\pi\indexRoot}{\numberOfRoots}},&\text{otherwise}\\
	\end{cases}~.
	\label{eq:schemeThree}
\end{align}
For instance, we obtain $\sum_{\indexMV=0}^{\numberOfMVs-1}\voteVectorEDBinaryEle[\indexED][\indexMV]2^\indexMV=0$ for $\voteVectorEDEle[\indexED][\indexMV]=-1$, $\forall\indexMV$,  as $\voteVectorEDBinaryEle[\indexED][\indexMV]=0$, $\forall\indexMV$. Hence, the radius of the $0$th root is $\radiusRoot^{-1}$, while the radius of any other root for $\indexRoot\neq0$ is set to $\radiusRoot$.
\begin{lemma}
	Let $\numberOfEdgeDevicesPlus[\indexMV]$ and $\numberOfEdgeDevicesMinus[\indexMV]$ denote the number of transmitters with positive and negative votes for $\indexMV$th MV computation.  For the mapping in \eqref{eq:schemeThree} with $\probability[{\voteVectorEDEle[\indexED][\indexMV']=1}]=\probability[{\voteVectorEDEle[\indexED][\indexMV']=-1}]=1/2$, $\forall\indexMV'$ and  $\indexMV'\neq\indexMV$,
	\begin{align}
		\expectationOperator[\left|{\polySeq[\superposedSeqP][ \radiusRoot\constante^{\frac{\constantj2\pi\indexRootRX}{\numberOfRoots}}]}\right|^2 ][] &= 	\frac{\numberOfEdgeDevicesPlus[\indexMV]\indicatorFunction[{\indexRootRXBin[\indexMV]=1}]+\numberOfEdgeDevicesMinus[\indexMV]\indicatorFunction[{\indexRootRXBin[\indexMV]=0}]}{2^{\log_2({\numberOfRoots})-1}}
		\funcSignalThree[\radiusRoot]
		\funcChannel[\radiusRoot]~,
		\label{eq:expectedPlusThree}
	\end{align}
	with 
	\begin{align}
		&\funcSignalThree[\radiusRoot]\triangleq	
\lastCorrTerm({\numberOfRoots+1})(\radiusRoot-\radiusRoot^{-1})^2\radiusRoot^{\numberOfRoots}
\numberOfRoots^2~,
	\end{align}	
	where the expectation in \eqref{eq:expectedPlusThree} is over the distribution of channels and votes  and $\indexRootRX = \sum_{\indexRootDigit=0}^{\numberOfMVs-1}\indexRootRXBin[\indexRootDigit]2^\indexRootDigit$ with $\indexRootRXBin[\indexRootDigit]\in\integers_2$, $\forall\indexRootDigit$.
	\label{lemma:exp3}
\end{lemma}
The proof is given in Appendix \ref{app:lemma3}.

Based on Lemma~\ref{lemma:exp3}, we can  obtain unbiased estimates of $\numberOfEdgeDevicesPlus[\indexMV]$ and $\numberOfEdgeDevicesMinus[\indexMV]$ as
\begin{align}
	\metricPlus[\indexMV] =	
	\frac{\sum_{\substack{\indexRootRX=0,\indexRootRXBin[\indexMV]=1}}^{\numberOfRoots-1}\left|{\polySeq[\receivedSeqP][ \radiusRoot\constante^{\frac{\constantj2\pi\indexRootRX}{\numberOfRoots}}]}\right|^2-\frac{\numberOfRoots}{2}\funcNoise[\radiusRoot]}{\funcSignalTwo[\radiusRoot]
		\funcChannel[\radiusRoot] 2^{-\log_2({\numberOfRoots})+1}
	}
	~,
\end{align}
and
\begin{align}
	\metricMinus[\indexMV]=	
	\frac{\sum_{\substack{\indexRootRX=0,\indexRootRXBin[\indexMV]=0}}^{\numberOfRoots-1}\left|{\polySeq[\receivedSeqP][ \radiusRoot\constante^{\frac{\constantj2\pi\indexRootRX}{\numberOfRoots}}]}\right|^2-\frac{\numberOfRoots}{2}\funcNoise[\radiusRoot]}{\funcSignalTwo[\radiusRoot]
	\funcChannel[\radiusRoot] 2^{-\log_2({\numberOfRoots})+1}
}
	~,
\end{align}
respectively, and derive the detector to obtain the  $\indexMV$th \ac{MV} as
\begin{align}
	\majorityVoteDetectedEle[\indexMV]=&\text{sign}\left(
			\sum_{\substack{\indexRootRX=0\\\indexRootRXBin[\indexMV]=1}}^{\numberOfRoots-1}\left|{\polySeq[\receivedSeqP][ \radiusRoot\constante^{\frac{\constantj2\pi\indexRootRX}{\numberOfRoots}}]}\right|^2
		-
		\sum_{\substack{\indexRootRX=0\\\indexRootRXBin[\indexMV]=0}}^{\numberOfRoots-1}\left|{\polySeq[\receivedSeqP][ \radiusRoot\constante^{\frac{\constantj2\pi\indexRootRX}{\numberOfRoots}}]}\right|^2
	\right).
	\label{eq:detector3}
\end{align}

Compared with Method~1 and Method~2, Method~3 uses $\numberOfRoots/2$ measurements for each test in \eqref{eq:detector3} to determine the \acp{MV}.  Hence, as demonstrated in Section~\ref{sec:numerical}, it yields a better \ac{CER}  at the expense of a lower computation rate, i.e., $\log_2(\numberOfRoots)$ \acp{MV} over $\numberOfRoots+\channelLength$ complex-valued resources. Also, the detector does not need the delay profile information. Note that Method~3 reduces to Method~2 for $\numberOfRoots=2$. 

\begin{figure*}[t]
	\centering
	\subfloat[Method 1 - Uncoded.]{\includegraphics[width=\textwidth/3]{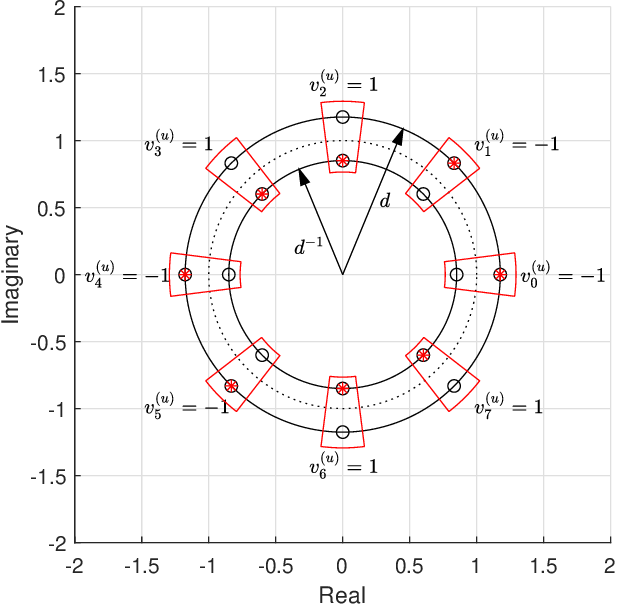}\label{subfig:example1}}
	\subfloat[Method 2 - Differential encoder.]{\includegraphics[width=\textwidth/3]{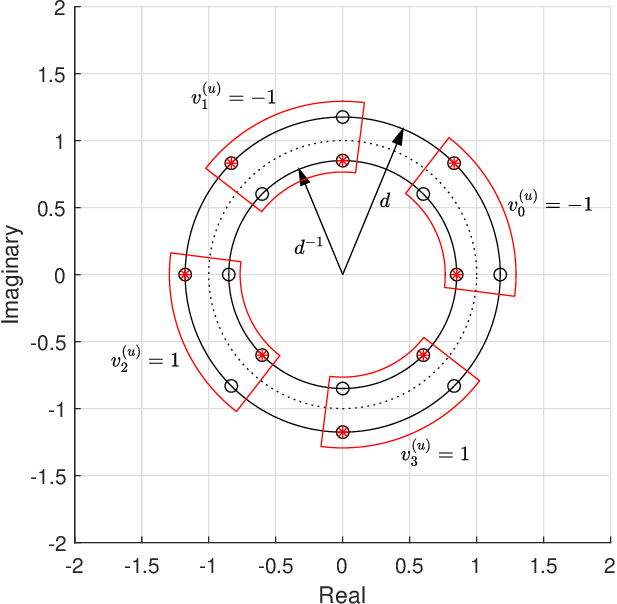}\label{subfig:example2}}		
	\subfloat[Method 3 - Index-based encoder.]{\includegraphics[width=\textwidth/3]{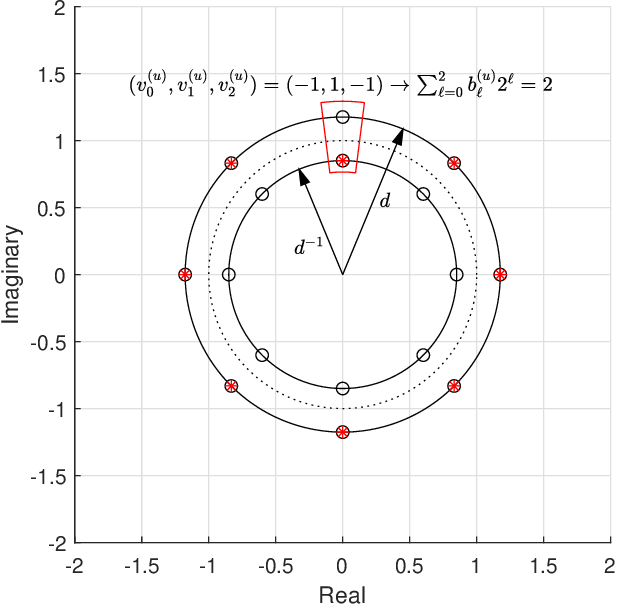}\label{subfig:example3}}
	\caption{Example zero placements for Methods~1-3. The star and circle markers indicate the chosen zeros and the possible zero locations for a  Huffman polynomial, respectively.}	
	\label{fig:examples}
\end{figure*}

In \figurename~\ref{fig:examples}, we exemplify the zero placements for Methods~1-3 and $\numberOfRoots=8$. In  \figurename~\ref{fig:examples}\subref{subfig:example1}, we assume that the votes at the transmitter are  $(\voteVectorEDEle[\indexED][0],\mydots,\voteVectorEDEle[\indexED][7])=(-1,-1,1,1,-1,-1,1,1)$. By following \eqref{eq:schemeOne}, the zeros  (see the points marked by stars in \figurename~\ref{fig:examples}\subref{subfig:example1}) are chosen based on the values of $\voteVectorEDEle[\indexED][\indexMV]$. In  \figurename~\ref{fig:examples}\subref{subfig:example2}, we consider Method~2, and the votes are  $(\voteVectorEDEle[\indexED][0],\voteVectorEDEle[\indexED][1],\voteVectorEDEle[\indexED][2],\voteVectorEDEle[\indexED][3])=(-1,-1,1,1)$. In this method,  two zeros are allocated for each vote, and the zeros alternate their radii based on the value of the vote by \eqref{eq:schemeTwo}. Finally, in  \figurename~\ref{fig:examples}\subref{subfig:example3}, we show the zero placement for Method~3 for $(\voteVectorEDEle[\indexED][0],\voteVectorEDEle[\indexED][1],\voteVectorEDEle[\indexED][2])=(-1,1,-1)$. Since we obtain   $\sum_{\indexMV=0}^{\numberOfMVs-1}\voteVectorEDBinaryEle[\indexED][\indexMV]2^\indexMV=2$ for $(\voteVectorEDBinaryEle[\indexED][0],\voteVectorEDBinaryEle[\indexED][1],\voteVectorEDBinaryEle[\indexED][2])=(0,1,0)$ from \eqref{eq:schemeThree}, the zero indexed by $\indexRoot=2$ changes its position while the other zeros remain on the circle with the radius $\radiusRoot$. {\color{\reviewColor}In \tablename~\ref{table:comp}, we summarize the pros and cons of the proposed methods.}

\newcolumntype{T}[1]{>{\centering\arraybackslash\hspace{0pt}}m{#1}} 
\begin{table}[]
	\color{\reviewColor}
	\caption{Comparison of methods.}	
	\centering
	\begin{tabular}{l|T{0.75in}|T{0.75in}|T{0.75in}|}
		%\cline{2-4}
		& Method 1                  & Method 2     & Method 3     \\ \hline
		CSI  & Not required                  & Not required & Not required \\ \hline
		PDP  & Required                  & Not required & Not required \\ \hline
		Rate& $\frac{\numberOfRoots}{\numberOfRoots+\channelLength}$ &      $\frac{\numberOfRoots/2}{\numberOfRoots+\channelLength}$        &       $\frac{\log_2(\numberOfRoots)}{\numberOfRoots+\channelLength}$       \\ \hline
		Reliability& Low &   Low           &    High          \\ \hline
	\end{tabular}
	\label{table:comp}
\end{table}	

\begin{figure*}[t]
	\centering
	\includegraphics[width = \textwidth]{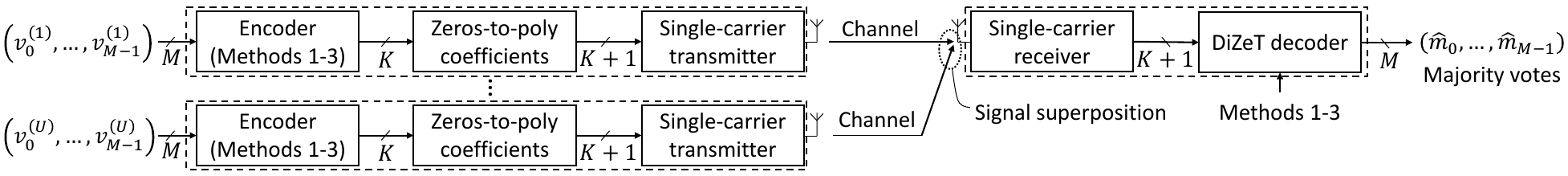}
	\caption{Transmitter and receiver block diagrams.}
	\label{fig:txrx}
\end{figure*}
Finally, we provide the transmitter and receiver block diagrams in \figurename~\ref{fig:txrx}. After the encoder generates the  zeros, i.e., a zero codeword of length $\numberOfRoots$, the zero codeword is converted to polynomial coefficients  of length $\numberOfRoots+1$.  After the conversion, the polynomial coefficients are transmitted with a single-carrier transmitter (e.g., upsampling and pulse shaping). We refer the reader to \cite{Sahin_2016} for the variants of a single-carrier waveform. The receiver receives the sum of the transmitted signals. After processing the superposed signal with a single-carrier receiver (e.g., matched filter and down-sampling), the receiver uses a \ac{dizet} decoder, i.e., \eqref{eq:detector1}, \eqref{eq:detector2}, or \eqref{eq:detector3}, to obtain the \acp{MV}. {\color{\reviewColor}It is worth noting that the proposed methods can also be utilized with \ac{OFDM} as long as the elements of the encoded sequence are mapped to multiple \ac{OFDM} symbols and subcarriers \textit{within} the coherence time and bandwidth (i.e., $\channelLength=1$ tap).}

{
\color{\reviewColor}	
\subsection{Complexity Analysis and Numerical Stability}
On the transmitter side, converting the encoded zeros to polynomial coefficients introduces an additional complexity to the radio.  In \cite[Eq. (8)]{Philipp_2019principles}, it is shown that the conversion can be handled with an iterative algorithm that incrementally builds the polynomial function by using each zero one step at a time.  The time complexity of this algorithm is $\mathcal{O}(\numberOfRoots^2)$. As noted in \cite{Philipp_2019principles}, MATLAB's $\mathtt{poly}$ function also uses the same expansion algorithm.\footnote{\color{\reviewColor}The version of MATLAB is 2024a.} It is also possible to reduce the complexity to $\mathcal{O}(\numberOfRoots\log^2(\numberOfRoots))$ by recursively grouping the zeros into halves and using the  fast Fourier transform for polynomial multiplication. 

Based on our analysis, the $\mathtt{poly}$ function does not provide numerically stable results for large $\numberOfRoots$ values. To address this issue, we use the implementation based on the derivation in Appendix~\ref{app:zeroToCoef}. In this method, we evaluate the polynomial function at  $\polyVariable=\constante^{\constantj2\pi\frac{\indexSample}{\numberOfRoots+1}}$
for $\indexSample\in\{0,1,\mydots,\numberOfRoots\}$ by using its zeros, and calculate the $(\numberOfRoots+1)$-point \ac{IDFT} of the resulting sequence to obtain the polynomial coefficients.
This algorithm  evaluates  \eqref{eq:keyeq} for $\numberOfRoots+1$ different values, each requiring  $\numberOfRoots$ multiplications. Thus, the time complexity of this method is also $\mathcal{O}(\numberOfRoots^2)$. However, it is more stable than the $\mathtt{poly}$ function as it does not accumulate the numerical errors through iterations.

On the receiver side, the complexity increases because of evaluating the polynomial $\polySeq[{\receivedSeqP}][\polyVariable]$ at $2\numberOfRoots$ zeros with \ac{dizet} decoders, i.e., \eqref{eq:detector1}, \eqref{eq:detector2}, and \eqref{eq:detector3}. The complexity of evaluating an $n$th order polynomial function is $\mathcal{O}(n)$ with Horner's method \cite{Horner_1819}. Thus, the additional time complexity due to a \ac{dizet} decoder can be obtained as $\mathcal{O}(\numberOfRoots^2)$. Note that the numerical stability of Horner's method can be improved further with state-of-the-art solutions (e.g., see \cite{Langlois_2007} and the references therein).
}

%In Appendix~\ref{app:zeroToCoef}, we provide how the zeros can be converted to the polynomial coefficients by using \ac{DFT}. Note that an iterative algorithm discussed in \cite[Eq. (8)]{Philipp_2019principles} can also be used for this conversion. 

\section{Computation-error rate Analysis}
\label{sec:performance}
For Methods~1-3, we can express the \ac{CER} for the $\indexMV$th \ac{MV} as
\begin{align}
	&\computationErrorRate[{\indexMV}] = \begin{cases}
		\probability[{\metricPlus[\indexMV]-\metricMinus[\indexMV]<0}]~, & \numberOfEdgeDevicesPlus[\indexMV]>\numberOfEdgeDevicesMinus[\indexMV]\\
		1-\probability[{\metricPlus[\indexMV]-\metricMinus[\indexMV]<0}]~, &
		\numberOfEdgeDevicesPlus[\indexMV]<\numberOfEdgeDevicesMinus[\indexMV]\\
		1~, & \numberOfEdgeDevicesPlus[\indexMV]=\numberOfEdgeDevicesMinus[\indexMV]
		\label{eq:cerDef}
	\end{cases}~,
\end{align}
where the third case in \eqref{eq:cerDef} is because $\sign({\metricPlus[\indexMV]-\metricMinus[\indexMV]})$ is almost surely not zero due to the noisy reception in communication channels. We can also express $\probability[{\metricPlus[\indexMV]-\metricMinus[\indexMV]<0}]$  as
\begin{align}
	\probability[{\metricPlus[\indexMV]-\metricMinus[\indexMV]<0}] &=\expectationOperator[{\CDF[{\metricPlus[\indexMV]-\metricMinus[\indexMV]}][][x;\voteAll]}][{\voteAllWithout}]\nonumber\\&=\frac{1}{2^{\numberOfMVs(\numberOfEdgeDevices-1)}}\sum_{\forall\voteAllWithout}{\CDF[{\metricPlus[\indexMV]-\metricMinus[\indexMV]}][][x;\voteAll]}]~,
	\label{eq:probDef}
\end{align}
where  $\CDF[{\metricPlus[\indexMV]-\metricMinus[\indexMV]}][][x;\voteAll]$ is the \ac{CDF} of  $\metricPlus[\indexMV]-\metricMinus[\indexMV]$ given all votes  $\voteAll\triangleq(\voteVectorAcrossED[\indexMV],\voteAllWithout)$ for $\voteVectorAcrossED[\indexMV]={(\oneVector[{\numberOfEdgeDevicesPlus[\indexMV]}],-\oneVector[{\numberOfEdgeDevicesMinus[\indexMV]}])}$ and $\voteAllWithout\triangleq(\voteVectorAcrossED[1],\mydots,\voteVectorAcrossED[\indexMV-1],\voteVectorAcrossED[\indexMV+1],\mydots,\voteVectorAcrossED[\numberOfEdgeDevices])$. Hence,  we need an analytical expression of $\CDF[{\metricPlus[\indexMV]-\metricMinus[\indexMV]}][][0;\voteAll]$ to obtain $\probability[{\metricPlus[\indexMV]-\metricMinus[\indexMV]<0}]$. To this end, we use  the following result from \cite{sahin2023reliable}:

\begin{lemma}[\cite{sahin2023reliable}]
	Let $\valOne[\indexRootRX]$ and $\valTwo[\indexRootRX]$ be independent exponential random variables with the rate $\rateOne[{\indexRootRX}]$ and $\rateTwo[{\indexRootRX}]$, respectively, $\forall\indexRootRX\in\{0,\mydots\numberOfRoots-1\}$. For $\metricOne=\sum_{\indexRootRX=0}^{\numberOfRoots-1}\valOne[\indexRootRX]$ and $\metricTwo=\sum_{\indexRootRX=0}^{\numberOfRoots-1}\valTwo[\indexRootRX]$, 	$\CDF[{\metricOne-\metricTwo}][][x]$ can be calculated as 
	\begin{align}
		\CDF[{\metricOne-\metricTwo}][][x]&=	\frac{1}{2}-\int_{-\infty}^{\infty}\frac{\charFcnOne[\integralVar]\charFcnTwoConj[\integralVar]}{2\pi\constantj\integralVar} \constante^{-\constantj\integralVar\CDFvariable}d\integralVar~,
		\label{eq:cdfMain}
	\end{align} where
	\begin{align}
		\charFcnOne[\integralVar]=\prod_{{\indexRootRX=0}}^{\numberOfRoots-1}\frac{1}{1-\constantj\integralVar{\rateOne[{\indexRootRX}]}^{-1}}~, 
	\end{align}
	and
	\begin{align}
		\charFcnTwo[\integralVar]=\prod_{{\indexRootRX=0}}^{\numberOfRoots-1}\frac{1}{1-\constantj\integralVar\rateTwo[{\indexRootRX}]^{-1}}~.
	\end{align}
	\label{lemma:errProb}
\end{lemma}
Lemma~\eqref{lemma:errProb} exploits the characteristic functions of the exponential distribution, convolution theorem, and the inversion formula given in \cite{waller_1995inversionCDF}. We can now calculate the $\CDF[{\metricPlus[\indexMV]-\metricMinus[\indexMV]}][][0;\voteAll]$ for Methods~1-3 theoretically as follows:
\begin{corollary}
	$\CDF[{\metricPlus[\indexMV]-\metricMinus[\indexMV]}][][0;\voteAll]$ for Method 1 can be calculated
by evaluating \eqref{eq:cdfMain} at $x=
{\funcNoise[\radiusRoot]}/({\funcSignalOne[\radiusRoot]
	\funcChannel[\radiusRoot]})-
{\funcNoise[\radiusRoot^{-1}]}/({\funcSignalOne[\radiusRoot^{-1}]
	\funcChannel[\radiusRoot^{-1}]})$ for
	\begin{align}
		\charFcnOne[\integralVar]=\frac{1}{1-\constantj\integralVar\rate[{+}]^{-1}}~, 
	\end{align}
	and
	\begin{align}
		\charFcnTwo[\integralVar]=\frac{1}{1-\constantj\integralVar\rate[{-}]^{-1}}~,
	\end{align}
	with 
	\begin{align}
		\rate[{+}]^{-1}=\frac{1}{\funcSignalOne[\radiusRoot]}\sum_{\indexED=1}^{\numberOfEdgeDevices}|\polySeq[\codedSeqP[\indexED]][{\radiusRoot\constante^{\frac{\constantj2\pi\indexRootRX}{\numberOfRoots}}}]|^2+\frac{\funcNoise[\radiusRoot]}{\funcSignalOne[\radiusRoot]\funcChannel[\radiusRoot]}~, \nonumber
	\end{align}
and
\begin{align}
		\rate[{-}]^{-1}=\frac{1}{\funcSignalOne[\radiusRoot^{-1}]}\sum_{\indexED=1}^{\numberOfEdgeDevices}|\polySeq[\codedSeqP[\indexED]][{\radiusRoot^{-1}\constante^{\frac{\constantj2\pi\indexRootRX}{\numberOfRoots}}}]|^2+\frac{\funcNoise[\radiusRoot^{-1}]}{\funcSignalOne[\radiusRoot^{-1}]\funcChannel[\radiusRoot^{-1}]}~. \nonumber
\end{align}
	\label{corr:cermethod1}
\end{corollary}
\begin{corollary}[]
$\CDF[{\metricPlus[\indexMV]-\metricMinus[\indexMV]}][][0;\voteAll]$ for Method 3 can be calculated by evaluating 
\eqref{eq:cdfMain} at $x=0$ with
	\begin{align}
		\charFcnOne[\integralVar]=\prod_{\substack{\indexRootRX=0\\\indexRootRXBin[\indexMV]=1}}^{\numberOfRoots-1}\frac{1}{1-\constantj\integralVar\rate[{\indexRootRX}]^{-1}}~, 
	\end{align}
	and
	\begin{align}
		\charFcnTwo[\integralVar]=\prod_{\substack{\indexRootRX=0\\\indexRootRXBin[\indexMV]=0}}\frac{1}{1-\constantj\integralVar\rate[{\indexRootRX}]^{-1}}~, 
	\end{align}
	for  $\rate[{\indexRootRX}]^{-1}=\funcChannel[\radiusRoot]\sum_{\indexED=1}^{\numberOfEdgeDevices}|\polySeq[\codedSeqP[\indexED]][{\radiusRoot\constante^{\frac{\constantj2\pi\indexRootRX}{\numberOfRoots}}}]|^2+\funcNoise[\radiusRoot]$.
	\label{corr:cerMethod32}
\end{corollary}
The proofs of Corollary~\ref{corr:cermethod1} and Corollary~\ref{corr:cerMethod32} are deferred to Appendix~\ref{app:cerCorr}.
\begin{corollary}
Since Method 3 reduces to Method 2 for $\numberOfRoots=2$, the CER for Method 2 can be also directly calculated by using Corollary~\eqref{corr:cerMethod32}. 
\label{corr:cerMethod3}
\end{corollary}

It is worth emphasizing that the integral in \eqref{eq:cdfMain} can be evaluated numerically for a given set of rate values for Methods 1-3. Also,  the sum in \eqref{eq:probDef} can be intractable for large $\numberOfEdgeDevices$ and $\numberOfMVs$. To address this issue, we calculate the average of the integral in \eqref{eq:cdfMain}  over a few realizations of $\voteAll$, as done in \cite{sahin2023reliable} in this study.

\section{Numerical Results}
\label{sec:numerical}
In this section, we assess the proposed methods numerically. We first generate the results on \ac{CER}, \ac{PMEPR}, and resource utilization per MV computation. We then apply the proposed methods to a specific application, i.e., distributed median computation.  
%We set the channel parameters as $\channelLength\in\{1,5\}$ with $\decayingParameter=1$ and set the number of roots as $\numberOfRoots\in\{8,16,32,64,128\}$. 
We also compare our results with Goldenbaum’s non-coherent OAC scheme in  \cite{Goldenbaum_2013tcom} {\color{\reviewColor}and the coherent OAC scheme in \cite{Guangxu_2021}, i.e., \ac{OBDA}}. 
For Goldenbaum’s method,  we map the votes $-1$ and $1$ to the symbols $0$ and $2$, respectively. Afterward, the square root of the symbol is multiplied with a unimodular random sequence of length $\goldenbaumLength$. We choose the phase of an element of unimodular sequence uniformly between 0 and $2\pi$. At the receiver, the norm-square of the aggregated sequences is calculated, and the calculated value is scaled with $f(x)=(x-\noiseVariance)/\goldenbaumLength-\numberOfEdgeDevices$. Finally, the sign of the scaled value is calculated to obtain the \ac{MV}. To make a fair comparison, we set $\goldenbaumLength$ to the nearest integer of
$(\numberOfRoots+1)/\log_2(\numberOfRoots)$, resulting in $\numberOfMVs=\log_2(\numberOfRoots)$ \acp{MV} over $\numberOfRoots+1$ resources as in Method~3, approximately. For instance, for $\numberOfRoots=32$, Method~3 computes $5$~MVs by using $33$~resources for Method 3. Hence,  $\goldenbaumLength$ is set to $7$ as $(\numberOfRoots+1)/\log_2(\numberOfRoots)\approx6.6$, and 5 MVs are computed over 35 resources. {\color{\reviewColor}For OBDA, we map the votes to \ac{BPSK} symbols and consider OFDM transmission (i.e., $\channelLength=1$). Unless otherwise stated, we use \ac{TCI} for OBDA, where  the truncation level is set to $0.2$ (i.e., the nodes do not transmit if $|\channelSeqEle[\indexED][0]|^2\le 0.2$). For comparisons, we also demonstrate the performance of OBDA under phase synchronization errors, where we model the phase error at each node with a uniform \acl{PDF}, i.e., $\uniformDistribution[-120^\circ][120^\circ]$. Note that the superposition is still largely constructive under this model as we use \ac{BPSK} symbols for OBDA.}

\begin{figure*}
	\centering
	\subfloat[\color{\reviewColor}$\channelLength=1$ and $\numberOfRoots=8$.]{\includegraphics[width=\textwidth/3]{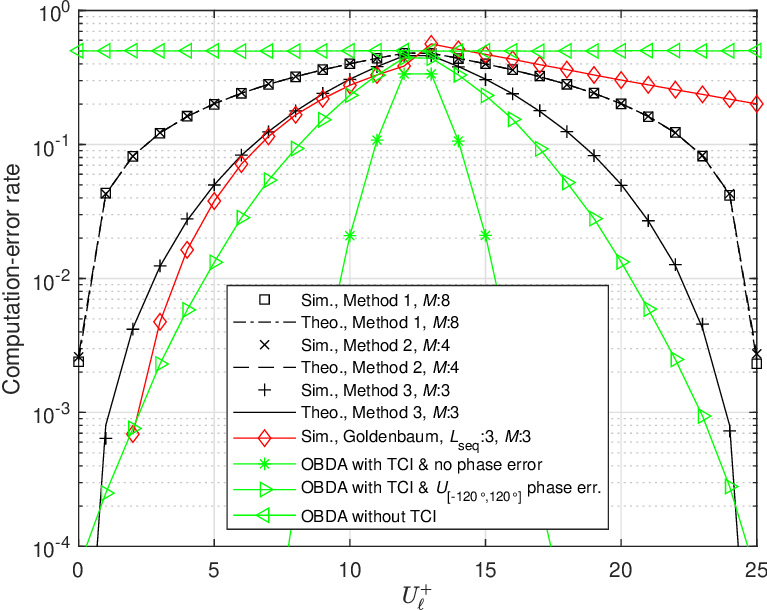}\label{subfig:CERL1K8}}
	\subfloat[\color{\reviewColor}$\channelLength=1$ and $\numberOfRoots=16$.]{\includegraphics[width=\textwidth/3]{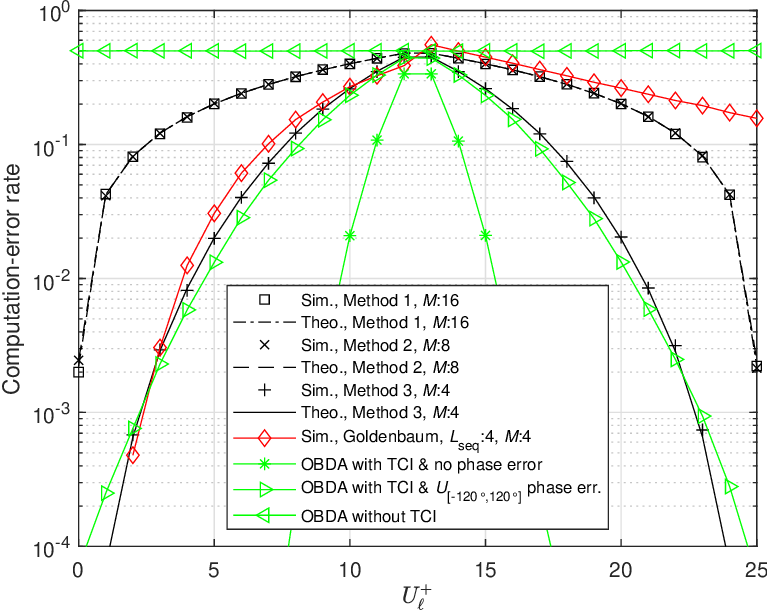}\label{subfig:CERL1K16}}		
	\subfloat[\color{\reviewColor}$\channelLength=1$ and $\numberOfRoots=32$.]{\includegraphics[width=\textwidth/3]{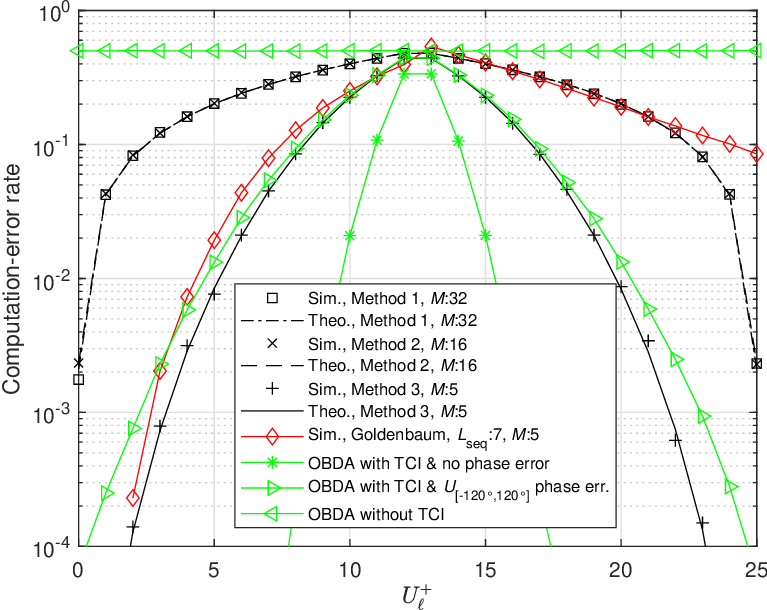}\label{subfig:CERL1K32}}		\\
	\subfloat[\color{\reviewColor}$\channelLength=5$, $\decayingParameter=1$, and $\numberOfRoots=8$.]{\includegraphics[width=\textwidth/3]{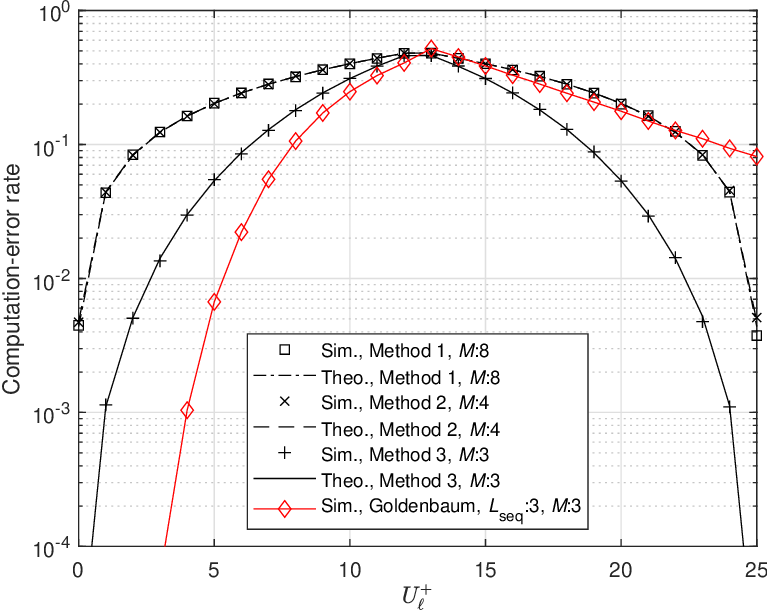}\label{subfig:CERL5K8}}
	\subfloat[\color{\reviewColor}$\channelLength=5$, $\decayingParameter=1$, and $\numberOfRoots=16$.]{\includegraphics[width=\textwidth/3]{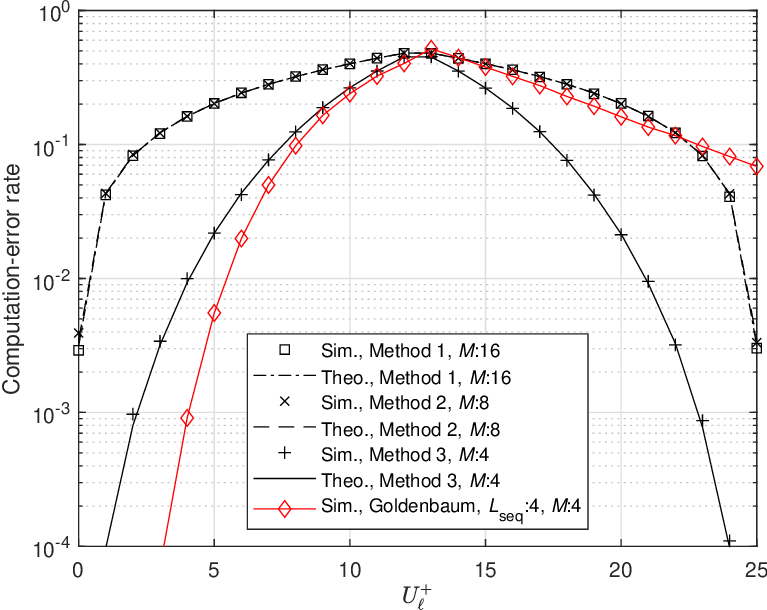}\label{subfig:CERL5K16}}		
	\subfloat[\color{\reviewColor}$\channelLength=5$, $\decayingParameter=1$, and $\numberOfRoots=32$.]{\includegraphics[width=\textwidth/3]{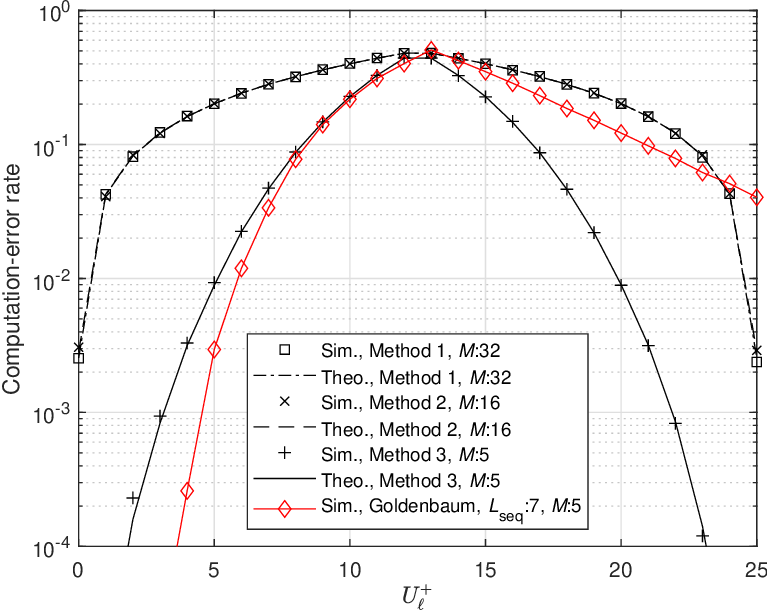}\label{subfig:CERL5K32}}		\\
	\caption{\color{\reviewColor}CER for a given $\numberOfEdgeDevicesPlus[\indexMV]$ ($\numberOfEdgeDevices=25$, $\SNR=10$~dB).}
	\label{fig:cerChannel}
\end{figure*}
In \figurename~\ref{fig:cerChannel}, we demonstrate the \ac{CER} performance of the schemes for a given number $\numberOfEdgeDevicesPlus[\indexMV]$ for $\numberOfEdgeDevices=25$ transmitters, $\SNR=10$~dB, and $\numberOfRoots\in\{8,16,32\}$. In \figurename~\ref{fig:cerChannel}\subref{subfig:CERL1K8}-\subref{subfig:CERL1K32}, we consider the case with $\channelLength=1$ tap. As can be seen from the results, increasing $|\numberOfEdgeDevicesPlus[\indexMV]-\numberOfEdgeDevicesMinus[\indexMV]|$ leads to a better \ac{CER} for all methods. This result is expected because the distance between two test values for the proposed methods with a DiZeT decoder increases with $|\numberOfEdgeDevicesPlus[\indexMV]-\numberOfEdgeDevicesMinus[\indexMV]|$. The performance of Method 1 and Method 2 are identical in \figurename~\ref{fig:cerChannel}\subref{subfig:CERL1K8}-\subref{subfig:CERL1K32} as we use the scalars in \eqref{eq:upOne} and \eqref{eq:umOne} for Method 1. However, Method 2 does not need a scalar, as seen in \eqref{eq:detector2}. Compared to Methods 1-2, Method~3 exploits redundancy and results in a remarkably better \ac{CER}, and the \ac{CER} improves further for increasing $\numberOfRoots$ at the expense of a reduced computation rate. For Goldenbaum's scheme, the transmitters do not transmit when the devices vote for $-1$, and the impact of the channel on the signal decreases for a smaller $\numberOfEdgeDevicesPlus[\indexMV]$. Thus, Goldenbaum's scheme causes an asymmetric behavior in CER in the range of $\numberOfEdgeDevicesPlus[\indexMV]$, and the CER degrades considerably by increasing for a large $\numberOfEdgeDevicesPlus[\indexMV]$. 
%We also observe that the CER is 1 for $\numberOfEdgeDevicesPlus[\indexMV]=8$ as the proposed detectors almost surely cannot detect $\numberOfEdgeDevicesPlus[\indexMV]=\numberOfEdgeDevicesMinus[\indexMV]$ under noisy reception.
{\color{\reviewColor}For OBDA, the CER performance is excellent when the nodes have perfect \ac{CSI}. However, under the phase synchronization errors, the CER increases quickly. Furthermore, if the TCI is not applied at the nodes due to the absence of CSI, OBDA cannot calculate the MV.}
 In \figurename~\ref{fig:cerChannel}\subref{subfig:CERL5K8}-\subref{subfig:CERL5K32}, we assess the \ac{CER} for a frequency-selective channel with $\channelLength=5$ taps and $\decayingParameter=1$. Compared to the results in \figurename~\ref{fig:cerChannel}\subref{subfig:CERL1K8}-\subref{subfig:CERL1K32}, the \ac{CER} slightly increases in the selective channel for all proposed methods, while it decreases for Goldenbaum's scheme. Nonetheless, it is still notable that a low-complexity DiZeT-based detector allows the receiver to compute the MVs without knowing the instantaneous CSI at the transmitters and receiver  (i.e., without phase and time synchronization across the devices).  Finally, the theoretical \ac{CER} results based on Corollaries~\ref{corr:cermethod1}-\ref{corr:cerMethod3} are well-aligned with the simulation results in \figurename~\ref{fig:cerChannel}.
 
 \begin{figure}[t]
	\centering
	\subfloat[\color{\reviewColor}$\channelLength=1$.]{\includegraphics[width=\figuresize]{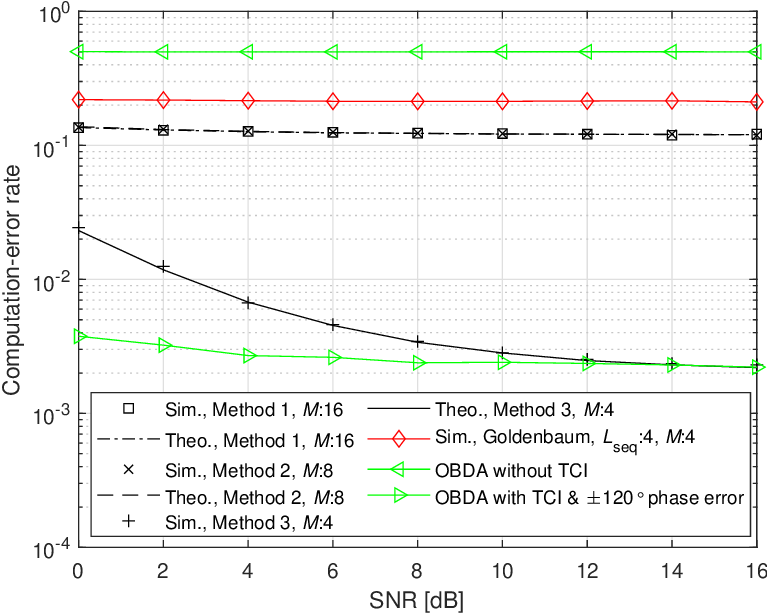}\label{subfig:snrcerL1}}\\
	\subfloat[\color{\reviewColor}$\channelLength=5$.]{\includegraphics[width=\figuresize]{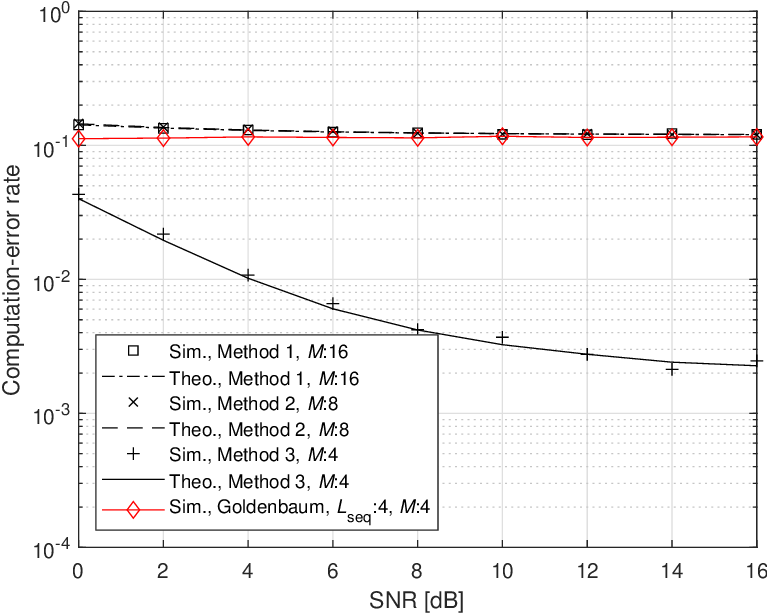}\label{subfig:snrcerL5}}		
	\caption{\color{\reviewColor}CER for a given SNR ($\numberOfEdgeDevices=25$, $\numberOfEdgeDevicesPlus[\indexMV]=22$, $\numberOfRoots=16$).}	
	\label{fig:snrcer}
\end{figure} 
{\color{\reviewColor}In \figurename~\ref{fig:snrcer}, we analyze how the \ac{SNR} changes the \ac{CER} for the same scenario in \figurename~\ref{fig:cerChannel}. For this analysis, we assume $\numberOfEdgeDevicesPlus[\indexMV]=22$ (i.e., the MV is 1) and $\numberOfRoots=16$, and sweep the SNR from 0 to 16~dB.  For all schemes, the CER curves explicitly reveal that increasing SNR does not improve the CER indefinitely. This implies that the distortion due to the fading channel is the main limiting factor for OAC, particularly for non-coherent schemes. Since Method 3 is more reliable than Methods 1-2, increasing SNR leads to a better CER performance. OBDA is more robust than the proposed methods but requires TCI and phase synchronization across the network. }

\begin{figure}[t]
	\centering
	\includegraphics[width = \figuresize]{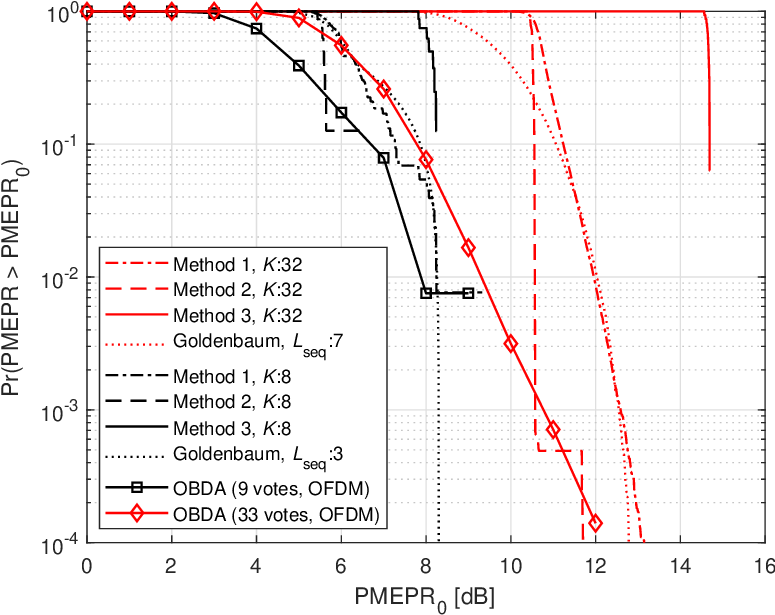}
	\caption{PMEPR distribution.}
	\label{fig:pmepr}
\end{figure}
In \figurename~\ref{fig:pmepr}, we analyze the \ac{PMEPR} distribution of the transmitted signals for $\numberOfRoots\in\{8,32\}$. For this analysis, we consider \ac{DFT-s-OFDM}, i.e., a variant of single-carrier waveform maintaining the linear convolution operation in \eqref{eq:receivedSamples} with zero-padding  \cite{Philipp_2021mumocz,Sahin_2016}, for our methods and Goldenbaum's scheme and OFDM for OBDA.   We set the \ac{DFT} and \ac{IDFT}  sizes to $\numberOfRoots+1$ and $16\times(\numberOfRoots+1)$, respectively, to over-sample the signal in the time domain by a factor of $16$. The results in \figurename~\ref{fig:pmepr} show that the instantaneous power of the transmitted signals can be high and Method~3 causes a  higher \ac{PMEPR} than Methods~1-2. The PMEPR distribution for Goldenbaum's scheme is also similar to Method~2. It is worth noting  that the PMEPR distribution of a single-carrier waveform depends on the distribution of the transmitted symbols. For the proposed methods, the elements of sequences originate from Huffman sequences. The magnitude of one of the sequence elements can be higher than that of the other elements in the sequence, leading to a high instantaneous signal power.  {\color{\reviewColor}For OBDA, the \ac{PMEPR} distribution is a function of the channel selectivity due to the TCI. In \figurename~\ref{fig:pmepr}, we assume a flat-fading channel for OBDA. While the likelihood of observing an OFDM symbol with a high \ac{PMEPR} increases with the number of subcarriers for OBDA, its PMEPR characteristics is better than the proposed methods and Goldenbaum's scheme with single-carrier waveform.
It is worth noting that the proposed methods can also be used with OFDM, as discussed in Section~III. For example, if the coherence bandwidth is sufficiently large, the encoded sequences can be mapped to contiguous OFDM subcarriers.  Under this mapping, the PMEPR of the transmitted signals  reduce to  1.54~dB and 1.79~dB for $\numberOfRoots=32$,  and $\numberOfRoots=8$, respectively, for all methods (not shown in \figurename~\ref{fig:pmepr}). This appealing result is expected as Huffman sequences have an identical \ac{AACF} and  their impulse-like \ac{AACF} result in a low PMEPR for OFDM transmission \cite{sahin_2020gm,davis_1999,Golay_1961}.}

\def\spectralEfficiency{\rho}
\begin{figure}
	\centering
	\includegraphics[width = \figuresize]{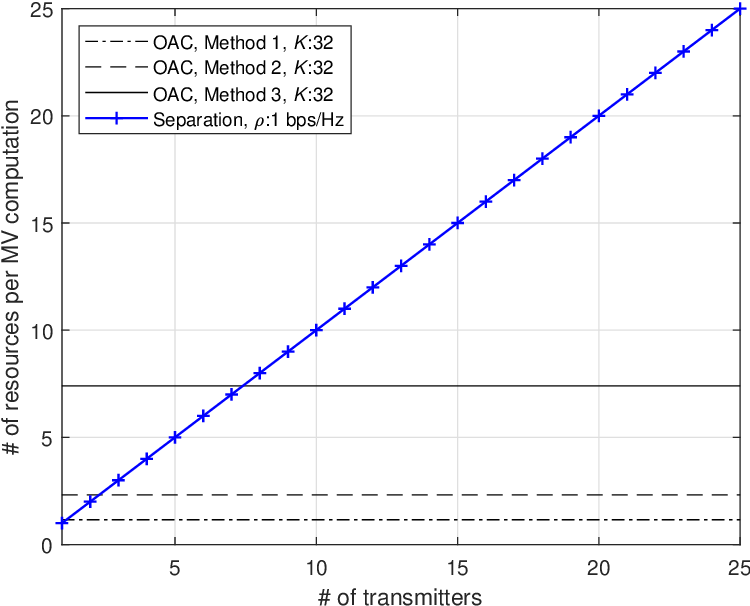}
	\caption{The number resources consumed per  MV computation for a given number of transmitters ($\channelLength=1$).}
	\label{fig:rate}
\end{figure}
In~\figurename~\ref{fig:rate}, we compare
Methods~1-3 with the case where the receiver computes the MV after it acquires each vote over orthogonal resources (i.e., separation of computation and communication) in terms of resource utilization. For the separation, we assume the spectral efficiency is $\spectralEfficiency=1$~bps/Hz. Since a vote can be represented with a single bit, we can compute the number of resources needed per MV computation as $\numberOfEdgeDevices/\spectralEfficiency$~resources. Based on the computation rates discussed in Section~\ref{sec:scheme}, the number of resources per MV are $(\numberOfRoots+\channelLength)/\numberOfRoots$, $(\numberOfRoots+\channelLength)/(\numberOfRoots/2)$, and $(\numberOfRoots+\channelLength)/(\log_2(\numberOfRoots)$ for Methods~1-3, respectively. In~\figurename~\ref{fig:rate}, we plot the number  of  resources consumed per MV computation for a given number of transmitters $\numberOfEdgeDevices$ for $\numberOfRoots=32$~zeros and $\channelLength=5$~taps. For Methods~1-3, the number of resources needed per MV can be calculated as $1.2$, $2.3$, and $7.4$ resources/MV, while it linearly increases with the number of transmitters for the separation. Method~3 becomes more efficient than the separation when at least $\numberOfEdgeDevices=8$ devices are in the network.

\def\indexIteration{i}
\def\parameterED[#1]{s_{\indexMV,#1}}
\def\medianTrue{c_{\indexMV}}
\def\medianEstimate[#1]{\hat{c}_{\indexMV}^{(#1)}}
\def\learningRate[#1]{\mu^{(#1)}}
\def\medianSweep{c}
\def\medianLoss[#1]{L_{\indexMV}(#1)}
\def\numberOfRealizations{N}
Next, we evaluate the schemes for  a distributed median computation scenario. For this application,
let $\parameterED[\indexED]$ denote the $\indexMV$th parameter at the $\indexED$th device,
and the goal is to compute the median value of the elements in $\{\parameterED[1],\mydots,\parameterED[\numberOfEdgeDevices]\}$ in a distributed manner, $\forall\indexMV\in\{0,1,\mydots,\numberOfMVs-1\}$. To this end, let us express the  median as a point minimizing the sum of distances to the parameters at the devices as
\begin{align}
	\medianTrue \triangleq \arg\min_{\medianSweep}\medianLoss[\medianSweep]~,
	\label{eq:problemMedian}
\end{align}
where $\medianTrue$ is the median value and $\medianLoss[\medianSweep]=\sum_{\indexED=1}^{\numberOfEdgeDevices}\norm{\medianSweep-\parameterED[\indexED]}_2$ is the corresponding loss function. Since $\medianLoss[\medianSweep]$ is convex, \eqref{eq:problemMedian} can be solved  iteratively as
\begin{align}
	\medianEstimate[\indexIteration+1]=&\medianEstimate[\indexIteration]-\learningRate[\indexIteration]\frac{d\medianLoss[\medianSweep]}{d\medianSweep}\bigg|_{\medianSweep=\medianEstimate[\indexIteration]}\nonumber\\
	=&\medianEstimate[\indexIteration]-\learningRate[\indexIteration]\sum_{\indexED=1}^{\numberOfEdgeDevices}\signNormal[{\medianEstimate[\indexIteration]-\parameterED[\indexED]}]~,
	\label{eq:medianGD}
\end{align}
where $\medianEstimate[\indexIteration]$ is an estimate of $\medianTrue$ and $\learningRate[\indexIteration]$ is the learning rate at the $\indexIteration$th iteration. Since the gradient direction can also be used for solving \eqref{eq:problemMedian} with an accuracy of $\pm\learningRate[\indexIteration]$, \eqref{eq:medianGD} can be modified  as
\begin{align}
	\medianEstimate[\indexIteration+1]=\medianEstimate[\indexIteration]-\learningRate[\indexIteration]\signNormal[{\sum_{\indexED=1}^{\numberOfEdgeDevices}\signNormal[{\medianEstimate[\indexIteration]-\parameterED[\indexED]}]}]~,
	\label{eq:medianMV}
\end{align}
where the update in \eqref{eq:medianMV} is well-aligned with the \ac{MV} computation problem in \eqref{eq:mvProblem}. In the case of a distributed scenario, the devices do not share their parameters in the network to promote privacy. Instead, the $\indexED$th device sets the $\indexMV$th vote as $\voteVectorEDEle[\indexED][\indexMV]=\signNormal[{\medianEstimate[\indexIteration]-\parameterED[\indexED]}]$ for the $\indexMV$th parameter at the $\indexIteration$th iteration, and all devices access the spectrum concurrently for \ac{OAC}. After the receiver computes the $\indexMV$th \ac{MV} with \ac{OAC} and updates $\medianEstimate[\indexIteration]$ as in \eqref{eq:medianMV}, it shares $\medianEstimate[\indexIteration+1]$ in the downlink for the next iteration.

\begin{figure*}
	\centering
	\subfloat[$\channelLength=1$ and $\numberOfRoots=8$.]{\includegraphics[width=\textwidth/3]{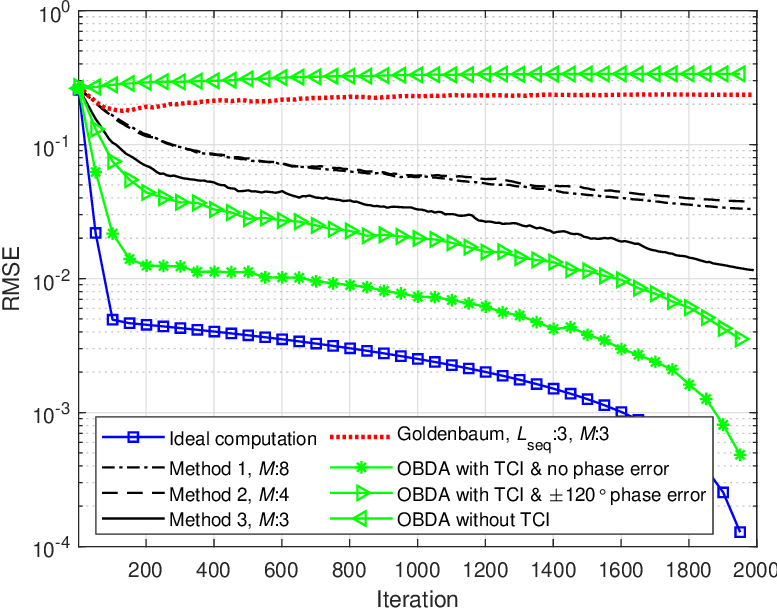}\label{subfig:MSEL1K8}}
	\subfloat[$\channelLength=1$ and $\numberOfRoots=32$.]{\includegraphics[width=\textwidth/3]{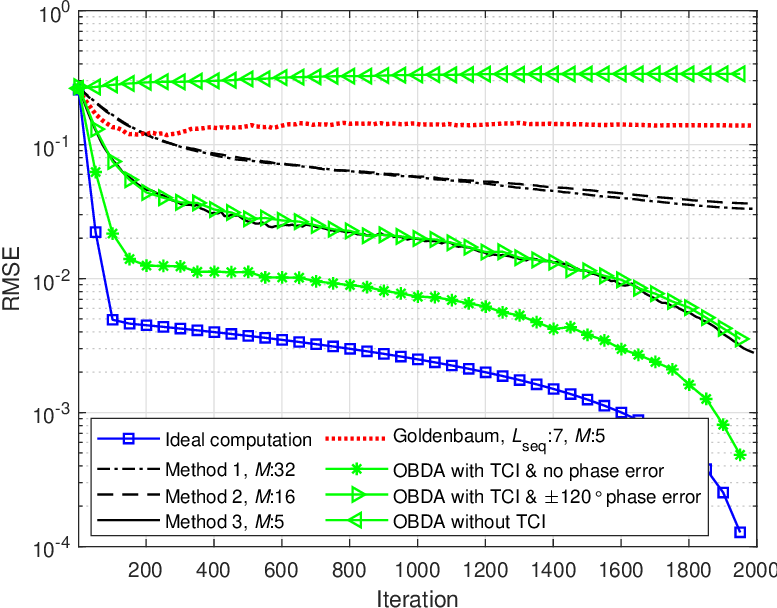}\label{subfig:MSEL1K32}}		
	\subfloat[$\channelLength=1$ and $\numberOfRoots=128$.]{\includegraphics[width=\textwidth/3]{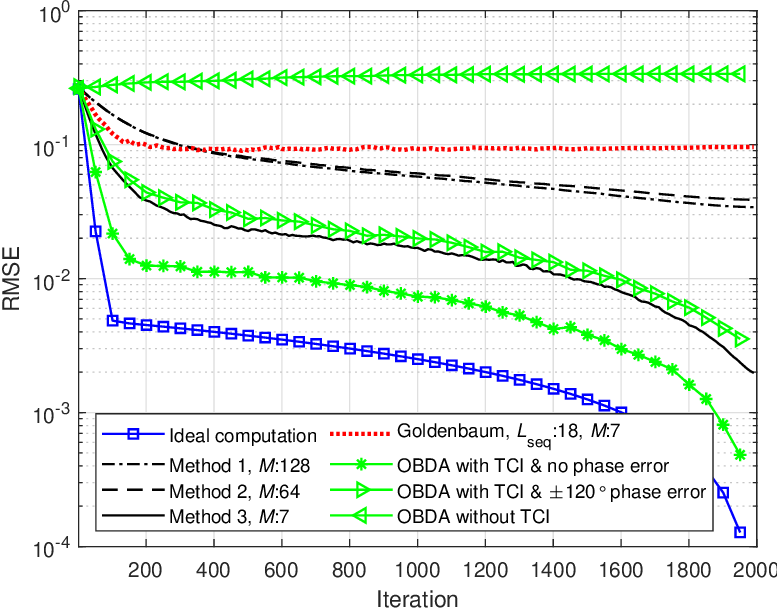}\label{subfig:MSEL1K128}}		\\
	\subfloat[$\channelLength=5$, $\decayingParameter=1$, and $\numberOfRoots=8$.]{\includegraphics[width=\textwidth/3]{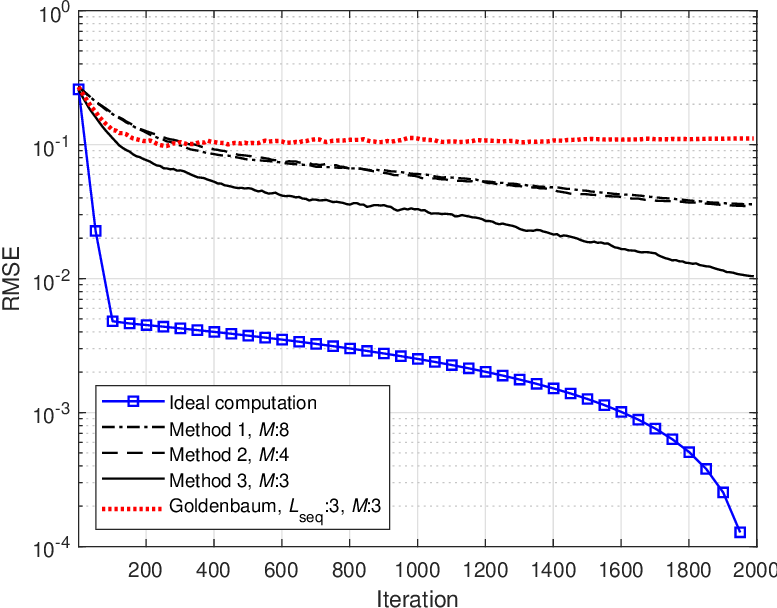}\label{subfig:MSEL5K8}}
	\subfloat[$\channelLength=5$, $\decayingParameter=1$, and $\numberOfRoots=32$.]{\includegraphics[width=\textwidth/3]{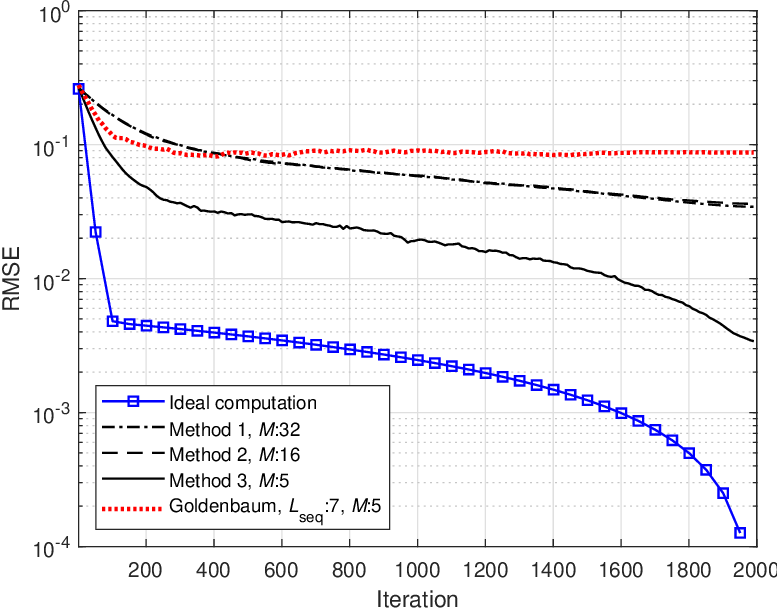}\label{subfig:MSEL5K32}}		
	\subfloat[$\channelLength=5$, $\decayingParameter=1$, and $\numberOfRoots=128$.]{\includegraphics[width=\textwidth/3]{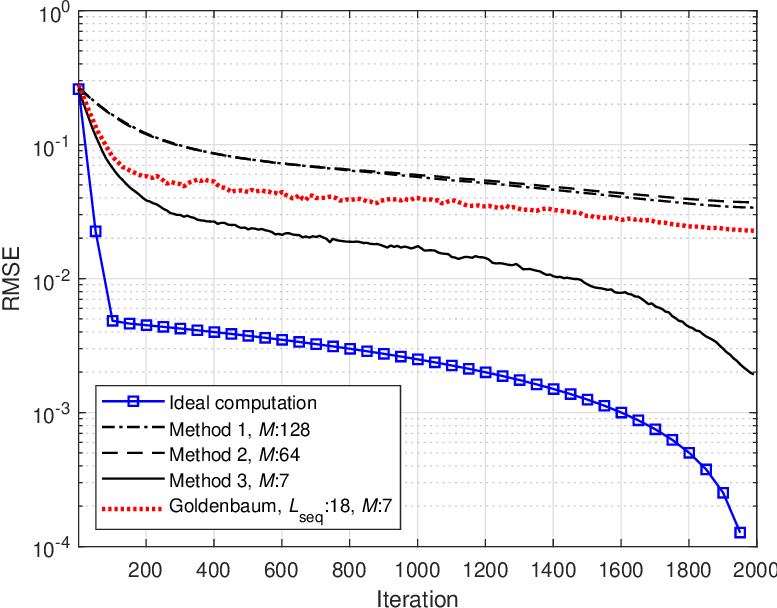}\label{subfig:MSEL5K128}}		\\
	\caption{RMSE for different number of roots and channel length ($\numberOfEdgeDevices=25$ transmitters, $\SNR=10$~dB).}
	\label{fig:rmseChannel}
\end{figure*}
In \figurename~\ref{fig:rmseChannel}, we plot the \ac{RMSE} of $\medianEstimate[\indexIteration]$, i.e., the square root of the arithmetic mean of $|{\medianEstimate[\indexIteration]-\medianTrue}|^2$, over the communication rounds   for $\numberOfRoots\in\{8,32,128\}$ over 1000 realizations. We consider  $\numberOfEdgeDevices=25$~devices and assume that $\parameterED[\indexED]\sim\uniformDistribution[-\sqrt{3}][\sqrt{3}]$, $\forall\indexED$, and reduce $\learningRate[\indexIteration]$ from 0.01 to 1e-5 linearly over the iterations. We also generate results when the MV computation occurs without impairments to provide a reference curve. In \figurename~\ref{fig:rmseChannel}\subref{subfig:MSEL1K8}-\subref{subfig:MSEL1K128} and \figurename~\ref{fig:rmseChannel}\subref{subfig:MSEL5K8}-\subref{subfig:MSEL5K128}, we consider the scenarios with $\channelLength=1$ and $\channelLength=5$ taps, respectively. Similar to the \ac{CER} results in \figurename~\ref{fig:cerChannel}, Method 3 is superior to Methods 1 and 2, while the performance of Method 1 and Method 2 are almost identical. Increasing the number of roots $\numberOfRoots$ also improves the performance of Method 3. For instance, the RMSE reduces to 0.002 for $\numberOfRoots=128$ from $0.01$ for $\numberOfRoots=8$.  Also, the impact of $\channelLength$ on the RMSE results is negligibly small for the proposed methods. Goldenbaum's scheme performs worse than Method 3 for $\goldenbaumLength=3$. However, its performance improves in more selective channels. A larger sequence length $\goldenbaumLength$ leads to a better result, as seen \figurename~\ref{fig:cerChannel}\subref{subfig:MSEL1K128}. This is because Goldenbaum's method is inherently sensitive to the cross-correlation of the sequences used at the transmitters, and the interference due to the cross-products decreases on average for larger sequence lengths. {\color{\reviewColor}As expected, OBDA is superior to the proposed methods and Goldenbaum's approach in an ideal scenario, i.e., when the perfect CSI is available. However, under phase synchronization errors, its performance degrades and is similar to that of Method~3. Also, we observe that OBDA does not work without TCI.}
 We observe a large gap between the ideal MV computation scenario and all methods based on OAC. Nonetheless, the OAC does not reveal the votes explicitly to the receiver due to the signal superposition while utilizing the spectrum efficiently through simultaneous transmissions on the same resources. In addition, the proposed OAC methods do not use the \ac{CSI} at the receiver or transmitters, i.e., reducing the potential overheads.

\section{Concluding Remarks}
\label{sec:conclusion}
In this work, we introduce a new strategy to compute the MV function over the air and discuss three different methods. Fundamentally, the proposed methods rely on nullifying a transmitter's contribution to the superposed value by encoding the votes, i.e., $+1$ and $-1$, into the zeros of a Huffman polynomial. We prove that this strategy non-coherently superposes the votes on two different test values, and a DiZeT decoder can be used for MV computation. The proposed methods inherently result in a trade-off between the computation rate, CER, and applicability. Method 1 has the highest computation rate. However, the decoder needs the PDP of the channel apriori. Method 2 improves Method 1's applicability to practice as the decoder does not require the PDP information by using a differential encoder. Method 3 is superior to Method 2 in terms of CER, but the computation rate is reduced further. We analyze the CER theoretically for all methods and provide analytical expressions that match the simulation results well. Finally, we show that the proposed methods can be applied to a distributed median computation scenario based on MV computation.

The proposed methods potentially lead to several interesting research directions. For instance, in this work, we choose the radii of roots to maximize the minimum distance between zeros as in \cite{Philipp_2019principles,Philipp_2020practical,Philipp_2021mumocz}. Also, we use two different radii for the roots based on Huffman polynomials. Hence, an unanswered challenge in this work is optimizing the number of radii and their radii for OAC. Secondly, in this work, we consider MV computation. How can we improve the methods for other nomographic functions? In this direction, the representation of integers in binary or balanced systems, as in \cite{sahin2022md}, may be explored. As we demonstrated, the PMEPR of the Huffman sequences for a single-carrier waveform can be high. Hence, another direction is reducing the PMEPR of the transmitted signals for the single-carrier waveform. Finally, assessing the performance of the proposed methods for training a neural network with federated learning is another angle that can be pursued.

\appendices
\section{Proof of Lemma~\ref{lemma:exp} }
\label{app:lemma1}
\begin{proof}[Proof of Lemma~\ref{lemma:exp}]
	We can express the right-hand side of \eqref{eq:expectedPlus} as
	\begin{align}
		&\expectationOperator[\left|{\polySeq[\superposedSeqP][ \radiusRoot\constante^{\frac{\constantj2\pi\indexMV}{\numberOfRoots}}]}\right|^2 ][]\nonumber\\
		&\stackrel{(a)}{=}\sum_{\indexED=1}^{\numberOfEdgeDevices}
		\expectationOperator[|{\polySeq[\channelSeqP[\indexED]][{
				\radiusRoot\constante^{\frac{\constantj2\pi\indexMV}{\numberOfRoots}}
			}]}|^2][]
		\expectationOperator[|{\polySeq[\codedSeqP[\indexED]][{
				\radiusRoot\constante^{\frac{\constantj2\pi\indexMV}{\numberOfRoots}}
			}]}|^2][]\nonumber\\
		&\stackrel{(b)}{=}\sum_{\substack{\indexED=1\\\voteVectorEDEle[\indexED][\indexMV]=1
		}}^{\numberOfEdgeDevices}
		\expectationOperator[|{\polySeq[\channelSeqP[\indexED]][{
				\radiusRoot\constante^{\frac{\constantj2\pi\indexMV}{\numberOfRoots}}
			}]}|^2][]
		\expectationOperator[|{\polySeq[\codedSeqP[\indexED]][{
				\radiusRoot\constante^{\frac{\constantj2\pi\indexMV}{\numberOfRoots}}
			}]}|^2][]~,
		\label{eq:expectedSuperpose}
	\end{align}	
	where (a) is because the channels and transmitted signals are independent random variables and (b) is because $\polySeq[\codedSeqP[\indexED]][ \radiusRoot\constante^{\frac{\constantj2\pi\indexMV}{\numberOfRoots}}]=0$ for any $\indexED$ with $\voteVectorEDEle[\indexED][\indexMV]=-1$. 
	By using \eqref{eq:normalization},
	\begin{align}
		&\expectationOperator[|{\polySeq[\codedSeqP[\indexED]][{
				\radiusRoot\constante^{\frac{\constantj2\pi\indexMV}{\numberOfRoots}}
			}]}|^2][]=	\expectationOperator[\Big|{\codedSeqEle[\indexED][\numberOfRoots]\prod_{\indexRoot=0}^{\numberOfRoots-1} \left(\radiusRoot\constante^{\frac{\constantj2\pi\indexMV}{\numberOfRoots}}-\rootCodedSeq[\indexED][\indexRoot]\right)}\Big|^2][]\nonumber\\
		&~~~~~~~~~~~~~=	\lastCorrTerm({\numberOfRoots+1})\expectationOperator[{\prod_{\indexRoot=0}^{\numberOfRoots-1} \frac{\Big|\radiusRoot\constante^{\frac{\constantj2\pi\indexMV}{\numberOfRoots}}-\rootCodedSeq[\indexED][\indexRoot]\Big|^2}{|\rootCodedSeq[\indexED][\indexRoot]|}}][]~.
		\label{eq:expectedTxSignalGeneral}
	\end{align}
	For any $\indexED$ with $\voteVectorEDEle[\indexED][\indexMV]=1$, we can  calculate
	\begin{align}
		&\expectationOperator[{|\polySeq[\codedSeqP[\indexED]][{
				\radiusRoot\constante^{\frac{\constantj2\pi\indexMV}{\numberOfRoots}}
			}]|^2| \voteVectorEDEle[\indexED][\indexMV]=1}][]
		\nonumber\\
%		&=		\lastCorrTerm({\numberOfRoots+1})\frac{(\radiusRoot-\radiusRoot^{-1})^2}{\radiusRoot^{-1}}\expectationOperator[{\prod_{\substack{\indexRoot=0\\\indexRoot\neq\indexMV}}^{\numberOfRoots-1} \frac{|\radiusRoot\constante^{\frac{\constantj2\pi\indexMV}{\numberOfRoots}}-\rootCodedSeq[\indexED][\indexRoot]|^2}{|\rootCodedSeq[\indexED][\indexRoot]|}}][]\nonumber\\
		&=	\lastCorrTerm({\numberOfRoots+1})\frac{(\radiusRoot-\radiusRoot^{-1})^2}{\radiusRoot^{-1}}\prod_{\substack{\indexRoot=0\\\indexRoot\neq\indexMV}}^{\numberOfRoots-1} \expectationOperator[{\frac{|\radiusRoot\constante^{\frac{\constantj2\pi\indexMV}{\numberOfRoots}}-\rootCodedSeq[\indexED][\indexRoot]|^2}{|\rootCodedSeq[\indexED][\indexRoot]|}}][]\nonumber\\
		&=	\lastCorrTerm({\numberOfRoots+1})\frac{(\radiusRoot-\radiusRoot^{-1})^2}{\radiusRoot^{-1}}\nonumber
		\\&~~~~
		\times\prod_{\substack{\indexRoot=1}}^{\numberOfRoots-1} {\frac{|\radiusRoot-\radiusRoot\constante^{\frac{\constantj2\pi\indexRoot}{\numberOfRoots}} |^2}{\radiusRoot}}\frac{1}{2} + {\frac{|\radiusRoot-\radiusRoot^{-1}\constante^{\frac{\constantj2\pi\indexRoot}{\numberOfRoots}} |^2}{\radiusRoot^{-1}}}\frac{1}{2} 
		\nonumber	\\
		&=	\lastCorrTerm({\numberOfRoots+1})(\radiusRoot-\radiusRoot^{-1})^2\radiusRoot^{{\numberOfRoots}}\nonumber
		\\&~~~~~~~
		\times\frac{1}{2^{\numberOfRoots-1}}\prod_{\substack{\indexRoot=1}}^{\numberOfRoots-1} {{|1-\constante^{\frac{\constantj2\pi\indexRoot}{\numberOfRoots}} |^2} + {|\radiusRoot-\radiusRoot^{-1}\constante^{\frac{\constantj2\pi\indexRoot}{\numberOfRoots}} |^2}}
		~.\label{eq:expectedSignal}
	\end{align}
	By plugging \eqref{eq:expectedChannel} and \eqref{eq:expectedSignal} in \eqref{eq:expectedSuperpose}, we obtain \eqref{eq:expectedPlus}. 
\end{proof}
\section{Proof of Lemma~\ref{lemma:exp2} }
\label{app:lemma2}
\begin{proof}[Proof of Lemma~\ref{lemma:exp2}]
	By using the same arguments for \eqref{eq:expectedSuperpose}, we can express the right-hand side of \eqref{eq:expectedPlusP1} as
	\begin{align}
		&\expectationOperator[\left|{\polySeq[\superposedSeqP][ \radiusRoot\constante^{\frac{\constantj2\pi2\indexMV}{\numberOfRoots}}]}\right|^2 ][]\nonumber\\
%		&\stackrel{(a)}{=}\sum_{\indexED=1}^{\numberOfEdgeDevices}
%		\expectationOperator[|{\polySeq[\channelSeqP[\indexED]][{
%				\radiusRoot\constante^{\frac{\constantj2\pi2\indexMV}{\numberOfRoots}}
%			}]}|^2][]
%		\expectationOperator[|{\polySeq[\codedSeqP[\indexED]][{
%				\radiusRoot\constante^{\frac{\constantj2\pi2\indexMV}{\numberOfRoots}}
%			}]}|^2][]\nonumber\\
		&=\sum_{\substack{\indexED=1\\\voteVectorEDEle[\indexED][\indexMV]=1
		}}^{\numberOfEdgeDevices}
		\expectationOperator[|{\polySeq[\channelSeqP[\indexED]][{
				\radiusRoot\constante^{\frac{\constantj2\pi2\indexMV}{\numberOfRoots}}
			}]}|^2][]
		\expectationOperator[|{\polySeq[\codedSeqP[\indexED]][{
				\radiusRoot\constante^{\frac{\constantj2\pi2\indexMV}{\numberOfRoots}}
			}]}|^2][]~.
		\label{eq:expectedSuperpose2}
	\end{align}	
	By using \eqref{eq:expectedTxSignalGeneral},
	for any $\indexED$ with $\voteVectorEDEle[\indexED][\indexMV]=1$, we can  calculate
	\begin{align}
		&\expectationOperator[{|\polySeq[\codedSeqP[\indexED]][{
				\radiusRoot\constante^{\frac{\constantj2\pi2\indexMV}{\numberOfRoots}}
			}]|^2| \voteVectorEDEle[\indexED][\indexMV]=1}][]
		\nonumber\\
		&=	\lastCorrTerm({\numberOfRoots+1})\frac{(\radiusRoot-\radiusRoot^{-1})^2}{\radiusRoot^{-1}}\frac{|\radiusRoot-\radiusRoot\constante^{\frac{2\pi}{\numberOfRoots}}|^2}{\radiusRoot}\nonumber\\
		&~~~~~~~~~~~
		\times\prod_{\substack{\indexRoot=0\\\indexRoot\neq\indexMV}}^{\frac{\numberOfRoots}{2}-1} \expectationOperator[{\frac{|\radiusRoot\constante^{\frac{\constantj2\pi\indexMV}{\numberOfRoots}}-\rootCodedSeq[\indexED][2\indexRoot]|^2}{|\rootCodedSeq[\indexED][2\indexRoot]|}
		\frac{|\radiusRoot\constante^{\frac{\constantj2\pi\indexMV}{\numberOfRoots}}-\rootCodedSeq[\indexED][2\indexRoot+1]|^2}{|\rootCodedSeq[\indexED][2\indexRoot+1]|}
		}][]\nonumber\\
		&=	\lastCorrTerm({\numberOfRoots+1})\frac{(\radiusRoot-\radiusRoot^{-1})^2}{\radiusRoot^{-1}}\frac{|\radiusRoot-\radiusRoot\constante^{\frac{2\pi}{\numberOfRoots}}|^2}{\radiusRoot}\nonumber
		\\&~~~~~~~~~~~~~~
		\times\prod_{\substack{\indexRoot=1}}^{\frac{\numberOfRoots}{2}-1} {\frac{|\radiusRoot-\radiusRoot\constante^{\frac{\constantj2\pi2\indexRoot}{\numberOfRoots}} |^2}{\radiusRoot}} {\frac{|\radiusRoot-\radiusRoot^{-1}\constante^{\frac{\constantj2\pi(2\indexRoot+1)}{\numberOfRoots}} |^2}{\radiusRoot^{-1}}}
		\frac{1}{2}
		\nonumber	\\
		&~~~~~~~~~~~~~~~~~~~~+ {\frac{|\radiusRoot-\radiusRoot^{-1}\constante^{\frac{\constantj2\pi2\indexRoot}{\numberOfRoots}} |^2}{\radiusRoot}} {\frac{|\radiusRoot-\radiusRoot\constante^{\frac{\constantj2\pi(2\indexRoot+1)}{\numberOfRoots}} |^2}{\radiusRoot^{-1}}}
		\frac{1}{2}
		\nonumber	\\
		&=	\lastCorrTerm({\numberOfRoots+1})(\radiusRoot-\radiusRoot^{-1})^2|1-\constante^{\frac{2\pi}{\numberOfRoots}}|^2\nonumber\radiusRoot^{\numberOfRoots}
		\\&~~~~~~
\times\frac{1}{2^{\frac{\numberOfRoots}{2}-1}}\prod_{\substack{\indexRoot=1}}^{\frac{\numberOfRoots}{2}-1} {{|1-\constante^{\frac{\constantj2\pi2\indexRoot}{\numberOfRoots}} |^2}} {{|\radiusRoot-\radiusRoot^{-1}\constante^{\frac{\constantj2\pi(2\indexRoot+1)}{\numberOfRoots}} |^2}}
\nonumber	\\
&~~~~~~~~~~~~~~~~~~~~+ 
{|1-\constante^{\frac{\constantj2\pi(2\indexRoot+1)}{\numberOfRoots}} |^2}
{{|\radiusRoot-\radiusRoot^{-1}\constante^{\frac{\constantj2\pi2\indexRoot}{\numberOfRoots}} |^2}} 
		~.\label{eq:expectedSignal2}
	\end{align}
	By plugging \eqref{eq:expectedChannel} and \eqref{eq:expectedSignal2} in \eqref{eq:expectedSuperpose2}, we obtain \eqref{eq:expectedPlusP1}.
	\end{proof}

\section{Proof of Lemma~\ref{lemma:exp3} }
\label{app:lemma3}
\begin{proof}[Proof of Lemma~\ref{lemma:exp3}]
We can express the right-hand side of \eqref{eq:expectedPlusThree} as
	\begin{align}
		&\expectationOperator[\left|{\polySeq[\superposedSeqP][ \radiusRoot\constante^{\frac{\constantj2\pi\indexRootRX}{\numberOfRoots}}]}\right|^2 ][]\nonumber\\
		&=\sum_{\substack{\indexED=1\\\sum_{\indexMV=0}^{\numberOfMVs-1}\voteVectorEDBinaryEle[\indexED][\indexMV]2^\indexMV=\indexRootRX
		}}^{\numberOfEdgeDevices}
		\expectationOperator[|{\polySeq[\channelSeqP[\indexED]][{
				\radiusRoot\constante^{\frac{\constantj2\pi\indexRootRX}{\numberOfRoots}}
			}]}|^2][]
		\expectationOperator[|{\polySeq[\codedSeqP[\indexED]][{
				\radiusRoot\constante^{\frac{\constantj2\pi\indexRootRX}{\numberOfRoots}}
			}]}|^2][]
		\nonumber\\
		&=
		\frac{\numberOfEdgeDevicesPlus[\indexMV]\indicatorFunction[{\indexRootRXBin[\indexMV]=1}]+\numberOfEdgeDevicesMinus[\indexMV]\indicatorFunction[{\indexRootRXBin[\indexMV]=0}]}{2^{\log_2({\numberOfRoots})-1}}
		\expectationOperator[|{\polySeq[\channelSeqP[\indexED]][{
				\radiusRoot\constante^{\frac{\constantj2\pi\indexRootRX}{\numberOfRoots}}
			}]}|^2][]\nonumber\\&~~~~~~~~~~~~~~~~~\times
		\expectationOperator[|{\polySeq[\codedSeqP[\indexED]][{
				\radiusRoot\constante^{\frac{\constantj2\pi\indexRootRX}{\numberOfRoots}}
			}]|^2|
		\sum_{\indexMV=0}^{\numberOfMVs-1}\voteVectorEDBinaryEle[\indexED][\indexMV]2^\indexMV=\indexRootRX
	}][]~.
		\label{eq:expectedSuperpose3}
	\end{align}	
	By using \eqref{eq:expectedTxSignalGeneral}, we can  calculate
	\begin{align}
		&\expectationOperator[|{\polySeq[\codedSeqP[\indexED]][{
				\radiusRoot\constante^{\frac{\constantj2\pi\indexRootRX}{\numberOfRoots}}
			}]|^2|
			\sum_{\indexMV=0}^{\numberOfMVs-1}\voteVectorEDBinaryEle[\indexED][\indexMV]2^\indexMV=\indexRootRX
		}][]
		\nonumber\\
		&=	\lastCorrTerm({\numberOfRoots+1})\frac{|\radiusRoot\constante^{\frac{\constantj2\pi\indexRootRX}{\numberOfRoots}}-\radiusRoot^{-1}\constante^{\frac{\constantj2\pi\indexRootRX}{\numberOfRoots}}|^2}{\radiusRoot^{-1}\constante^{\frac{\constantj2\pi\indexRootRX}{\numberOfRoots}}}
		\prod_{\substack{\indexRoot=0\\\indexRoot\neq\indexMV}}^{\numberOfRoots-1} {\frac{|\radiusRoot\constante^{\frac{\constantj2\pi\indexRootRX}{\numberOfRoots}}-\radiusRoot\constante^{\frac{\constantj2\pi\indexRoot}{\numberOfRoots}}|^2}{|\radiusRoot\constante^{\frac{\constantj2\pi\indexRoot}{\numberOfRoots}}|}
		}\nonumber\\
		&=	\lastCorrTerm({\numberOfRoots+1})(\radiusRoot-\radiusRoot^{-1})^2\radiusRoot^{\numberOfRoots}
		\prod_{\substack{\indexRoot=1}}^{\numberOfRoots-1} {{|1-\constante^{\frac{\constantj2\pi\indexRoot}{\numberOfRoots}} |^2}}
		\nonumber\\
		&=	\lastCorrTerm({\numberOfRoots+1})(\radiusRoot-\radiusRoot^{-1})^2\radiusRoot^{\numberOfRoots}
		\numberOfRoots^2
		~.\label{eq:expectedSignal3}		
	\end{align}
	By plugging \eqref{eq:expectedChannel} and \eqref{eq:expectedSignal3} in \eqref{eq:expectedSuperpose3}, we obtain \eqref{eq:expectedPlusThree}. 
\end{proof}

\section{From zeros to polynomial coefficients}
\label{app:zeroToCoef}

Consider the polynomial given in \eqref{eq:polyDef}. Let us define $\samplePoint[\indexSample][\indexED]$ as
\begin{align}
	\samplePoint[\indexSample][\indexED] \triangleq \polySeq[\codedSeqP[\indexED]][\polyVariable]|_{\polyVariable=\constante^{\constantj2\pi\frac{\indexSample}{\numberOfRoots+1}}}.
\end{align}
for $\indexSample\in\{0,1,\mydots,\numberOfRoots\}$. Thus, 
\begin{align}
	\samplePoint[\indexSample][\indexED]=\sum_{\indexTime=0}^{\numberOfRoots}\codedSeqEle[\indexED][\indexTime]\constante^{\constantj2\pi\frac{\indexSample\indexTime}{\numberOfRoots+1}}=
	\codedSeqEle[\indexED][\numberOfRoots]\prod_{\indexRoot=0}^{\numberOfRoots-1} (\constante^{\constantj2\pi\frac{\indexSample}{\numberOfRoots+1}}-\rootCodedSeq[\indexED][\indexRoot])
	~.
	\label{eq:keyeq}
\end{align}
The left-hand side of \eqref{eq:keyeq} corresponds to a $(K+1)$-point \ac{IDFT} of the sequence $\codedSeq[\indexED]=(\codedSeqEle[\indexED][0],\codedSeqEle[\indexED][1],\dots, \codedSeqEle[\indexED][\numberOfRoots])$. Hence, to obtain $\codedSeq[\indexED]$ from the zeros, we apply the $(K+1)$-point \ac{DFT} to the right-hand side of  \eqref{eq:keyeq} as
\begin{align}
	\codedSeqEle[\indexED][\indexTime]&=\frac{1}{\numberOfRoots+1}\sum_{\indexSample=0}^{\numberOfRoots}	\samplePoint[\indexSample][\indexED]\constante^{-\constantj2\pi\frac{\indexSample\indexTime}{\numberOfRoots+1}}\nonumber 
	\\
	&=\frac{\codedSeqEle[\indexED][\numberOfRoots]}{\numberOfRoots+1}\sum_{\indexSample=0}^{\numberOfRoots}\constante^{-\constantj2\pi\frac{\indexSample\indexTime}{\numberOfRoots+1}} \prod_{\indexRoot=0}^{\numberOfRoots-1} (\constante^{\constantj2\pi\frac{\indexSample}{\numberOfRoots+1}}-\rootCodedSeq[\indexED][\indexRoot])
\end{align}
for $\indexTime\in\{0,1,\mydots,\numberOfRoots\}$.

\section{Proofs of Corollary~\eqref{corr:cermethod1} and Corollary~\eqref{corr:cerMethod32}}
\label{app:cerCorr}
	\begin{proof}[Proof of Corollary~\eqref{corr:cermethod1}]
	By using \eqref{eq:upOne} and \eqref{eq:umOne}, we obtain 
	\begin{align}
		&\probability[{\metricPlus[\indexMV]-\metricMinus[\indexMV]<0};\voteAll]\nonumber\\&~~~~~~=
			\probability[\frac{\left|{\polySeq[\receivedSeqP][ \radiusRoot\constante^{\frac{\constantj2\pi\indexMV}{\numberOfRoots}}]}\right|^2}{\funcSignalOne[\radiusRoot]
			\funcChannel[\radiusRoot]}-	\frac{\left|{\polySeq[\receivedSeqP][ \radiusRoot^{-1}\constante^{\frac{\constantj2\pi\indexMV}{\numberOfRoots}}]}\right|^2}{\funcSignalOne[\radiusRoot^{-1}]
			\funcChannel[\radiusRoot^{-1}]}<x;\voteAll]\nonumber
	\end{align}
	for $x=
	{\funcNoise[\radiusRoot]}/({\funcSignalOne[\radiusRoot]
		\funcChannel[\radiusRoot]})-
	{\funcNoise[\radiusRoot^{-1}]}/({\funcSignalOne[\radiusRoot^{-1}]
		\funcChannel[\radiusRoot^{-1}]})$. For a given a set of votes, $|{\polySeq[\receivedSeqP][ \radiusRoot\constante^{\frac{\constantj2\pi\indexRootRX}{\numberOfRoots}}]}|^2/(\funcSignalOne[\radiusRoot]
		\funcChannel[\radiusRoot])$ is an exponential random variable with the mean $\rate[{+}]^{-1}$ because $\polySeq[\channelSeqP[\indexED]][{\radiusRoot\constante^{\frac{\constantj2\pi\indexRootRX}{\numberOfRoots}}}]$, $\forall \indexED$, and $\polySeq[\noiseSeqP][{\radiusRoot\constante^{\frac{\constantj2\pi\indexRootRX}{\numberOfRoots}}}]$ are independent random variables following zero-mean symmetric complex Gaussian distribution. A similar deduction can be made for $|{\polySeq[\receivedSeqP][ \radiusRoot^{-1}\constante^{\frac{\constantj2\pi\indexRootRX}{\numberOfRoots}}]}|^2/(\funcSignalOne[\radiusRoot^{-1}]
		\funcChannel[\radiusRoot^{-1}])$, leading to $\rate[{-}]^{-1}$.
\end{proof}
\begin{proof}[Proof of Corollary~\eqref{corr:cerMethod32}]
For a given a set of votes, $|{\polySeq[\receivedSeqP][ \radiusRoot\constante^{\frac{\constantj2\pi\indexRootRX}{\numberOfRoots}}]}|^2$ is an exponential random variable where its mean is $\rate[{\indexRootRX}]^{-1}=\funcChannel[\radiusRoot]\sum_{\indexED=1}^{\numberOfEdgeDevices}|\polySeq[\codedSeqP[\indexED]][{\radiusRoot\constante^{\frac{\constantj2\pi\indexRootRX}{\numberOfRoots}}}]|^2+\funcNoise[\radiusRoot]$ because  $\polySeq[\channelSeqP[\indexED]][{\radiusRoot\constante^{\frac{\constantj2\pi\indexRootRX}{\numberOfRoots}}}]$, $\forall \indexED$, and $\polySeq[\noiseSeqP][{\radiusRoot\constante^{\frac{\constantj2\pi\indexRootRX}{\numberOfRoots}}}]$ are independent random variables following zero-mean symmetric complex Gaussian distribution. 
\end{proof}

\bibliographystyle{IEEEtran}
\bibliography{references}

\end{document}